\documentclass[symmetry,article,submit,moreauthors,pdftex]{Definitions/mdpi}
\renewcommand{\linenumbers}{}
\usepackage{rotating}
\usepackage{braket}
\usepackage[section]{placeins}
\newcommand{\Lagr}{\mathcal{L}}

\firstpage{1}
\pubvolume{xx}
\issuenum{1}
\articlenumber{5}
\pubyear{2020}
\copyrightyear{2020}
\history{Received: date; Accepted: date; Published: date}

\Title{The ${\bar K}N$ interaction in higher partial waves}

\Author{Albert Feijoo$^{1,2}$\orcidB{}, Daniel Gazda$^2$\orcidD{}, Volodymyr Magas$^{3}$\orcidC{} and \`Angels Ramos$^{3}$\orcidA{} }

\AuthorNames{Albert Feijoo, Daniel Gazda, Volodymyr Magas and \`Angels Ramos}

\address{%
$^{1}$ \quad Instituto de F\'isica Corpuscular (centro mixto CSIC-UV), Institutos de Investigaci\'on de Paterna, Apartados 22085, 46071, Valencia, Spain\\ 
$^{2}$ \quad Nuclear Physics Institute of the Czech Academy of Sciences, 25068 \v{R}e\v{z}, Czech Republic\\
$^{3}$ \quad Departament de F\'isica Qu\`antica i Astrof\'isica and Institut de Ci\`encies del Cosmos (ICCUB), Universitat de Barcelona, Mart\'i i Franqu\`es 1, 08028 Barcelona, Spain \\
}

\abstract{We present a chiral ${\bar K}N$ interaction model that has
  been developed and optimized in order to account for the
  experimental data of inelastic ${\bar K}N$ reaction channels that
  open at higher energies. In particular, we study the effect of the
  higher partial waves which originate directly from the chiral
  Lagrangian, as they could supersede the role of high-spin resonances
  employed in earlier phenomenological models to describe
  meson--baryon cross sections in the 2~GeV region. We present a
  detailed derivation of the partial wave amplitudes that emerge from
  the chiral SU(3) meson-baryon Lagrangian up to the d-waves and
  next-to-leading order in the chiral expansion. We implement a
  nonperturbative unitarization in coupled channels and optimize the
  model parameters to a large pool of experimental data in the
  relevant energy range where these new contributions are expected to
  be important. The obtained results are encouraging. They indicate
  the ability of the chiral higher partial waves to extend the
  description of the scattering data to higher energies and to account
  for structures in the reaction cross sections that cannot be
  accommodated by theoretical models limited to the s-waves.}


\begin{document}


\section{Introduction}

Over the last forty years, strong interaction processes in the low-energy domain have been described with great success by chiral perturbation theory ($\chi$PT), an effective field theory with hadron degrees of freedom which respects all the
symmetries of the underlying theory of the strong interaction, Quantum Chromodynamics (QCD), particularly its spontaneously broken chiral symmetry~\cite{Gasser:1983yg}.

In the early years of $\chi$PT many studies were focused on the low-energy pion interactions with pions and nuclei, although extensions to the strangeness sector, invoking flavour SU(3) symmetry, were also explored. In particular, the ${\bar K}N$ interaction was especially challenging because the description of the scattering amplitude cannot be properly achieved within a perturbative treatment, owing to the presence of a resonance, the $\Lambda(1405)$, located 27~MeV below the ${\bar K}N$ threshold.

Another challenge in hadron physics over the past few decades has been to disentangle the nature of some hadron resonances that do not fit well in a conventional quark model description, according to which
baryons are composed by three quarks and mesons by a quark-antiquark pair~\cite{Guo:2017jvc}.
For many years, a large amount of theoretical and experimental activity has been devoted to the discovery and characterization of exotic mesons and baryons. One of the early evidences was provided, precisely, by the $\Lambda(1405)$ resonance, whose mass was systematically predicted to be too high by quark models but it found a better explanation if it was described as a $\bar K N$ quasi-bound state. This was already pointed out in the late fifties~\cite{Dalitz:1959dn,Dalitz:1960du}, and was later corroborated by models that built the meson--baryon interaction from a chiral effective Lagrangian, derived respecting the symmetries of QCD, and implementing unitarization~\cite{Kaiser:1995eg,Oset:1997it,Oller:2000fj,Lutz:2001yb,GarciaRecio:2002td,Borasoy:2005ie}.
The confirmation of the $\Lambda(1405)$ being essentially a
meson-baryon bound state came after establishing its double-pole nature~\cite{Oller:2000fj,Jido:2003cb} from comparing different experimental line shapes~\cite{Magas:2005vu}, which unambiguously indicates the different coupling strength to the meson-baryon components of the $\Lambda(1405)$ wave-function. See the recent overview on this issue in Ref.~\cite{Meissner:2020khl}.

Several experiments aiming at establishing the shape of the $\Lambda(1405)$, as those employing $pp$ reactions by the COSY~\cite{Zychor:2007gf} and HADES~\cite{HADES:2012csk} collaborations
or photo-{} or electro-production processes at LEPS~\cite{Niiyama:2008rt}
and CLAS~\cite{CLAS:2013rjt,CLAS:2013rxx,CLAS:2013zie}, together with the precise determination of the energy shift and width of the 1s
state in kaonic hydrogen measured by
SIDDHARTA~\cite{SIDD}, triggered a renewed interest in improving the chiral unitary theories~\cite{GO,IHW,Roca:2013cca,Mai:2014xna,Cieply:2016jby,Feijoo:2015yja, Ramos:2016odk, Feijoo:2018den} for a better description of the ${\bar K}N$ interaction and related phenomenology. In our previous works~\cite{Feijoo:2015yja, Ramos:2016odk, Feijoo:2018den}, the relevance of the terms next in the hierarchy after the lowest order Weinberg--Tomozawa (WT) term was studied, in connection with the inclusion of higher-energy experimental data. More explicitly, we first focused on the $K \Xi$ production reactions because they do not proceed directly via the lowest order WT contribution~\cite{Feijoo:2015yja}. Our further studies~\cite{Ramos:2016odk, Feijoo:2018den} indeed demonstrated that the so-called Born diagrams as well as the next-to-leading order (NLO) terms are far from being mere corrections when it comes to the reproduction of the $K^- p\to \eta\Lambda, \eta\Sigma^0, K^+\Xi^-, K^0\Xi^0$ reaction cross sections. In particular, the inclusion of isospin filtering reactions in~\cite{Feijoo:2018den,Feijoo:2015cca} served to emphasize their relevance for avoiding potential ambiguities in the isospin components of the scattering amplitude. All this translates into stronger constraints on the models by means of which one can derive more reliable values of the low-energy constants of the chiral Lagrangian.

In the same spirit, the inclusion of new ingredients that are expected to be specially relevant at higher energies could reveal more information about the physics behind the NLO terms of the chiral Lagrangian. In the present work, we explore the relevance of including partial waves higher than the $L=0$, which is usually the only component considered in the literature to study the ${\bar K}N$ scattering phenomenology. In particular, we focus on the p-wave contribution, the effect of which is expected to be non-negligible, as we aim at obtaining the ${\bar K}N$ scattering amplitudes at higher energies, necessary to describe the $\eta\Lambda$, $\eta\Sigma^0$, $K^0\Xi^0$ and $K^+\Xi^-$ production reactions. Extending the ${\bar K}N$ interaction to p-wave components is also relevant for studies of bound ${\bar K}$ mesons in nuclei~\cite{Garcia-Recio:2000seh,Cieply:2011yz,Hrtankova:2017zxw,Hrtankova:2019jky}, since their local momentum can acquire sizable values. Finally, from a formal perspective, it is interesting to explore whether the unitarized interaction in higher partial waves might give rise to dynamically generated states, similarly to the case of the $\Lambda(1405)$ resonance in s-wave.

Former studies of the effect of p-wave contributions derived from chiral Lagrangians have been done in the strangeness $S=0$ sector~\cite{CaroRamon:1999jf} and, more recently, in the $S=+1$ sector~\cite{Aoki:2018wug}. As for the $S=-1$ sector considered in the present work, the study of Ref.~\cite{Jido:2002zk}, limited to the lowest-order chiral Lagrangian, found the traditional low-energy data not to be sensitive to the new p-wave components. These findings were corroborated by a later study~\cite{Cieply:2015pwa} that followed similar ideas. The recent work of Ref.~\cite{Sadasivan:2018jig} obtains the s-{} and p-wave scattering amplitudes from the leading-order (LO) and NLO Lagrangians, the parameters of which are fitted to the low-energy data, also including the invariant $\pi\Sigma$ mass distributions from photo-production reactions at CLAS~\cite{CLAS:2013rjt}. A dynamically generated p-wave state with isospin one and $J^P=1/2^+$ is found in that study, which in fact mimics the absence of the $\Sigma^*(1385)$ resonance that plays a relevant role in the $\pi\Sigma$ distributions.

In the present work we consider the s-{} and p-wave contributions of the scattering amplitudes but, differently to Ref.~\cite{Sadasivan:2018jig}, we also employ the data of inelastic reactions opening at higher energies to constrain our models. This implies that the NLO Lagrangian, which we take from
Ref.~\cite{Aoki:2018wug}, must also consider terms that have usually been disregarded owing to their negligible effect at lower energies. We will show that the new terms of the NLO Lagrangian are relevant, being in fact strongly intertwined with the p-wave components of the ${\bar K}N$ interaction.

Our paper is organized as follows. In Sec.~\ref{sec:formalism} we develop the chiral unitary formalism, show the LO and NLO contributions of the chiral Lagrangian, give the explicit expressions of the corresponding interaction kernels---with their complete momentum dependence structure---and, finally, describe the procedure that allows one to obtain the partial-wave components of the scattering amplitude. The models studied in this work are presented in Sec.~\ref{sec:modelsData}, together with the data employed in the fits. Sec.~\ref{sec:results} is devoted to the discussion of our results, with a special focus on the role played by the p-wave amplitudes and the new terms of the NLO Lagrangian. A few concluding remarks are given in Sec.~\ref{sec:conclusions}.

\section{Formalism}
\label{sec:formalism}
Chiral unitary approaches (UChPT) have shown to be a powerful tool to treat the meson--baryon scattering at energies around resonances. These nonperturbative schemes prevent plain chiral perturbation theory from non converging and guarantee the unitarity and analyticity of the scattering amplitude. In the present work, unitarity is implemented
by solving
the Bethe--Salpeter~(BS) equation
with coupled channels. Following~\cite{Oset:1997it,Hyodo:2011ur}, the interaction kernel is conveniently split into its on-shell contribution and the corresponding off-shell one. The off-shell part gives rise to a tadpole-type diagram which can be
reabsorbed
into renormalization of couplings and masses and, hence, be omitted from the calculation. This
method
permits factorizing the interaction kernel and the scattering amplitude out of the integral equation, transforming a complex system of coupled integral equations into a simple system of algebraic equations which, in matrix form, reads:
\begin{equation}
T_{ij} ={(1-V_{il}G_l)}^{-1}V_{lj} ,
 \label{T_algebraic}
\end{equation}
where $V_{ij}$ is the driving kernel derived from the chiral Lagrangian, $T_{ij}$ is the corresponding scattering amplitude for the transition
from an \(i\) channel to a \(j\) one,
and $G_l$ is the loop function of the intermediate channel $l$, which reads:
\begin{equation} \label{Loop_integral}
G_l={\rm i}\int \frac{d^4q_l}{{(2\pi)}^4}\frac{2M_l}{{(P-q_l)}^2-M_l^2+{\rm i}\epsilon}\frac{1}{q_l^2-m_l^2+{\rm i}\epsilon} ,
\end{equation}
with $M_l$ and $m_l$ being the baryon and meson masses of the channel. As this function diverges logarithmically, a dimensional regularization scheme is applied which leads to the expression:
%
%
\begin{equation}
G_l = \frac{2M_l}{(4\pi)^2} \Bigg \lbrace a_l(\mu)+\ln\frac{M_l^2}{\mu^2}+\frac{m_l^2-M_l^2+s}{2s}\ln\frac{m_l^2}{M_l^2} +
\frac{q_{\rm cm}}{\sqrt{s}}\ln\left[\frac{(s+2\sqrt{s}q_{\rm cm})^2-(M_l^2-m_l^2)^2}{(s-2\sqrt{s}q_{\rm cm})^2-(M_l^2-m_l^2)^2}\right]\Bigg \rbrace .
 \label{dim_reg}
\end{equation}
The subtraction constants $a_l$ replace the divergence for a given dimensional regularization scale $\mu$, which is taken to be $1$~GeV in the present work. These constants are unknown parameters to be fitted to the experimental data. For the $S=-1$ meson--baryon interaction they amount to ten, one per channel, although it is quite common to reduce them to six taking into account isospin symmetry arguments, as we shall also do.

The kernel employed in the chiral unitary approaches is derived from
the SU(3) effective chiral Lagrangian, which provides the fundamental blocks of the interaction that preserve the symmetries of QCD employing hadron fields as
the relevant
degrees of freedom. Following a power counting scheme, these constituent pieces are arranged in the expansion by order of relevance (see for instance~\cite{Scherer:2002tk} for a more detailed explanation). At leading order (LO), the most general
Lagrangian
can be expressed as:
\begin{equation}
\Lagr_{\phi B}^{(1)}  =  i \langle \bar{B} \gamma_{\mu} [D^{\mu},B] \rangle
                            - M_0 \langle \bar{B}B \rangle
                            - \frac{1}{2} D \langle \bar{B} \gamma_{\mu}
                             \gamma_5 \{u^{\mu},B\} \rangle \\
                       - \frac{1}{2} F \langle \bar{B} \gamma_{\mu}
                               \gamma_5 [u^{\mu},B] \rangle,
\label{LagrphiB1}
\end{equation}
where $M_0$ is the common baryon octet mass in the chiral limit, the constants $D$, $F$ denote the axial vector couplings of the baryons to the mesons, and the symbol $\langle \cdot \rangle$ stands for the trace in flavour space. The baryon octet field ($N, \Lambda, \Sigma, \Xi $) is denoted by $B$, while the pseudoscalar meson octet field $\phi$ ($\pi, K, \eta$) enters in $u_\mu = i u^\dagger \partial_\mu U u^\dagger$, where $U(\phi) = u^2(\phi) = \exp{\left( \sqrt{2} i \phi/f \right)}$ (see~\cite{Feijoo:2015yja}) with $f$ being the pseudoscalar decay constant that acts as typifying scale factor in the expansion in powers of momentum. Finally, $[D_{\mu},B]$ contains the covariant derivative that accounts for the local character of the chiral transformation of $u$ and it is defined as:
\begin{equation}
[D_{\mu},B] = \partial_\mu B+ [\Gamma_{\mu},B] \,
\label{Cov_deriv}
\end{equation}
with $\Gamma_{\mu}= [u^\dagger , \partial_\mu u]/2$ being the chiral connection.
The next-to-leading order (NLO) contributions are given by:
\begin{equation}
\label{LagrphiB2}
\begin{split}
  \Lagr_{\phi B}^{(2)} &= b_D \langle \bar{B} \{\chi_+,B\} \rangle
  + b_F \langle \bar{B} [\chi_+,B] \rangle
  + b_0 \langle \bar{B} B \rangle \langle \chi_+ \rangle
  + d_1 \langle \bar{B} \{u_{\mu},[u^{\mu},B]\} \rangle \\
  &+ d_2 \langle \bar{B} [u_{\mu},[u^{\mu},B]] \rangle
  + d_3 \langle \bar{B} u_{\mu} \rangle \langle u^{\mu} B \rangle
  + d_4 \langle \bar{B} B \rangle \langle u^{\mu} u_{\mu} \rangle \\
  &- \frac{g_1}{8M_N^2} \langle \bar{B} \{u_{\mu},[u_{\nu}, \{D^{\mu},D^{\nu}\}B] \} \rangle
  - \frac{g_2}{8M_N^2} \langle \bar{B} [u_{\mu},[u_{\nu}, \{D^{\mu},D^{\nu}\}B] ] \rangle \\
  &- \frac{g_3}{8M_N^2} \langle \bar{B} u_{\mu} \rangle \langle [u_{\nu}, \{D^{\mu},D^{\nu}\}B] \rangle
  - \frac{g_4}{8M_N^2} \langle \bar{B} \{D^{\mu},D^{\nu}\} B \rangle \langle u_{\mu} u_{\nu} \rangle \\
  &- \frac{h_1}{4} \langle \bar{B} [\gamma^{\mu},\gamma^{\nu}] B u_{\mu} u_{\nu} \rangle
  - \frac{h_2}{4} \langle \bar{B} [\gamma^{\mu},\gamma^{\nu}] u_{\mu} [u_{\nu},B] \rangle
  - \frac{h_3}{4} \langle \bar{B} [\gamma^{\mu},\gamma^{\nu}] u_{\mu} \{u_{\nu},B\} \rangle \\
  &- \frac{h_4}{4} \langle \bar{B} [\gamma^{\mu},\gamma^{\nu}] u_{\mu} \rangle \langle u_{\nu},B \rangle + \text{h.c.},
\end{split}
\end{equation}
where $M_N$ stands for the nucleon mass, while the coefficients $b_D$, $b_F$, $b_0$, $d_i$ $(i=1,\ldots,4)$, $g_i$ $(i=1,\ldots,4)$ and $h_i$ $(i=1,\ldots,4)$
are
the low-energy constants (LECs) at
this order. Despite the symmetries of the underlying theory cannot fix these constants, some of them can be constrained by the mass splitting of baryons, the pion-nucleon sigma term and from the strangeness content of the proton~\cite{Gasser:1990ce}, and some others using data from low-energy meson--baryon interactions, such as the isospin even $\pi N$ s-wave scattering length~\cite{Koch:1985bn} and the isospin zero kaon--nucleon s-wave scattering length~\cite{Dover:1982zh}. However, we treat them as free parameters in the fitting procedure as it is usually done in the literature. The reason lies in the coupled-channel unitarization that produces amplitudes that go beyond tree level. The quantity $\chi_+ = 2 B_0 (u^\dagger \mathcal{M} u^\dagger + u \mathcal{M} u)$ explicitly breaks chiral symmetry via the quark mass matrix $\mathcal{M} = {\rm diag}(m_u, m_d, m_s)$, while $B_0 = - \braket{0 \vert \bar{q} q \vert 0} / f^2$ relates to the order parameter of spontaneously broken chiral symmetry.

Before proceeding further, we would like to make a few remarks on the NLO Lagrangian of Eq.~(\ref{LagrphiB2}), taken from~\cite{Aoki:2018wug}, where the authors study the $KN$ scattering amplitude in a chiral unitary approach looking for a possible broad resonance in the $S=+1$ sector. Formally speaking, the expression differs from the one we employed in our previous works~\cite{Feijoo:2015yja, Ramos:2016odk, Feijoo:2018den} in the terms with coefficients $g_i$ $(i=1,\ldots,4)$ and $h_i$ $(i=1,\ldots,4)$ (from now on \(g\)-terms and \(h\)-terms), which naturally appear from an extension of the SU(2) chiral Lagrangian introduced in~\cite{Bernard:1995dp,Fettes:2000gb,Lutz:2001yb}, but have been commonly discarded in the literature studying $S=-1$ meson--baryon scattering within the UChPT framework owing to their non significant role at low energies. Nevertheless, as our models explore higher energies, it is reasonable to consider these additional contributions. We will show that they
play a relevant role mainly in the $\bar{K}N$ transitions to the $\eta\Lambda$, $\eta\Sigma^0$, and $K \Xi$ channels. We also note that the $b_D$, $b_F$, $b_0$, $d_i$ and \(h\)-terms are invariant under the hermitian conjugation (h.c.) transformation in Eq.~(\ref{LagrphiB2}), hence their h.c.\ contributions can be reabsorbed in the LECs of the corresponding original terms. In contrast, the h.c.\ of the \(g\)-terms produces new structures in the Lagrangian, which have to be considered explicitly. Specially remarkable is the cancellation of the $g_3$ monomial, which implies the reduction of the parameter space in the derived models. Finally, it should be pointed out that other
versions for SU(3) Lagrangians at $O(q^2)$ are presented in~\cite{Frink:2004ic,Oller:2006yh}.

\begin{figure}[H]
\centering
\includegraphics[width=0.7\textwidth]{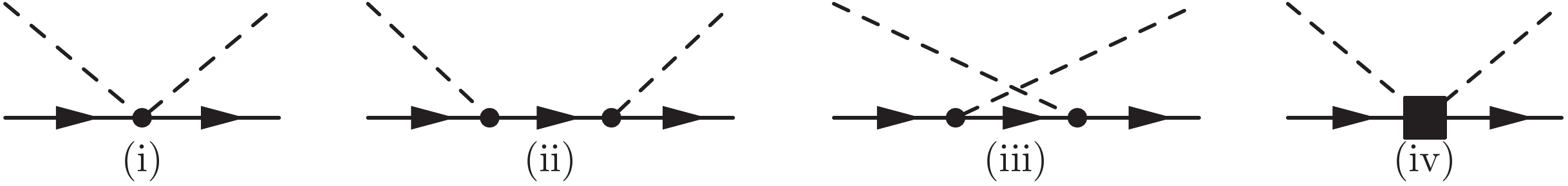}
\caption{\label{diagrams} Diagrammatic representation of the meson-baryon interaction kernels: Weinberg--Tomozawa term (i), direct and crossed Born terms (ii) and (iii), and NLO terms (iv). Dashed (solid) lines represent the pseudoscalar octet mesons (octet baryons).}
\end{figure}

Not without
a tedious calculation, one can derive the amplitudes for the $\phi_i B_i^s \to \phi_j B_j^{s'}$ processes (with
the incoming and outgoing spins $s,s'$, respectively) which are schematically represented in Fig.~\ref{diagrams}. Diagram (i) corresponds to the Weinberg--Tomozawa (WT) term that can be written in momentum space as:
\begin{equation}
\label{Kernel_WT}
 V_{ij}^{WT} =   -\frac{ N_i N_j}{4f^2} C_{ij} \left\lbrace (2\sqrt{s}{-}M_i{-}M_j)\chi_j^{\dagger s'} \chi_i^s +  \frac{2\sqrt{s}{+}M_i{+}M_j}{(E_i{+}M_i)(E_j{+}M_j)} \chi_j^{\dagger s'} \left[\vec{q}_j \cdot \vec{q}_i  + i (\vec{q}_j \times \vec{q}_i )\cdot \vec{\sigma} \right] \chi_i^s \right\rbrace .
\end{equation}
The indices $(i,j)$ cover all possible initial and final channels in the $S=-1, Q=0$ sector. The matrix of coefficients $C_{ij}$ can be found in Table VII of Ref.~\cite{Feijoo:2015yja}. The normalization factor $N$ is defined as $N=\sqrt{(M+E)/(2M)}$, with $M$ and $E$ being, respectively, the mass and energy of the baryon in the corresponding channel.
The two-component Pauli spinor of a baryon with spin projection $s$ is denoted by $\chi^s$. The symbol for the spin, only present in the spinors, should not be confused with $\sqrt{s}$ that represents the total energy of the meson-baryon system in the center-of-mass (CM) frame. The CM three-momentum of the meson (baryon) is given by $\vec{q}$ ($-\vec{q}$).

In Fig.~\ref{diagrams}, the Born contributions are represented by diagrams (ii) (direct Born term) and (iii) (crossed Born term), the vertices of which are obtained from the $D$ and $F$ terms of Eq.~(\ref{LagrphiB1}). The analytical form of the direct Born term is given by:
\begin{equation}
\label{Kernel_BD}
\begin{split}
  V_{ij}^{D} &= \frac{ N_iN_j}{12f^2} \sum_{k} \frac{C^{(\text{Born})}_{\bar{i}i,k}  C^{(\text{Born})}_{\bar{j}j,k}}{s-M_k^2}  \left\lbrace \vphantom{\frac{1}{(E_i+M_i)(E_j+M_j)}} (\sqrt{s}-M_k)(s+M_iM_j-\sqrt{s}(M_i+M_j))  \chi_j^{\dagger s'} \chi_i^s \right. \\
  &\left. +  \frac{(s+\sqrt{s}(M_i+M_j)+M_iM_j)(\sqrt{s}+M_k)}{(E_i+M_i)(E_j+M_j)} \chi_j^{\dagger s'} \left[ \vec{q}_j \cdot \vec{q}_i  + i (\vec{q}_j \times \vec{q}_i ) \cdot \vec{\sigma} \right] \chi_i^s  \right\rbrace
\end{split}
\end{equation}
and, similarly, the crossed Born term reads:
\begin{equation}
  \begin{split}
    V^{C}_{ij}  &= -\frac{ N_i N_j}{12f^2} \sum_{k} \frac{C^{(\text{Born})}_{\bar{j}k,i} C^{(\text{Born})}_{\bar{i}k,j}}{u-M_k^2}
    \left\lbrace
      \vphantom{\frac{\vec{q}_j \cdot \vec{q}_i  + i (\vec{q}_j \times \vec{q}_i ) \cdot \vec{\sigma}}{(E_i+M_i)(E_j+M_j)}}
      \left[
        \vphantom{M_i^2}
        u(\sqrt{s}+M_k)+\sqrt{s}(M_j(M_i+M_k)+M_iM_k)  \right.  \right. \\
    & \left.
      -M_j(M_i+M_k)(M_i+M_j) - M_i^2M_k \right]  \chi_j^{\dagger s'} \chi_i^s + \left[ u(\sqrt{s}-M_k) + \sqrt{s}(M_j(M_i+M_k)+M_iM_k)  \right. \\
      &\left. \left. + M_j(M_i+M_k)(M_i+M_j) + M_i^2M_k \right]  \chi_j^{\dagger s'} \frac{\vec{q}_j \cdot \vec{q}_i  + i (\vec{q}_j \times \vec{q}_i ) \cdot \vec{\sigma}}{(E_i+M_i)(E_j+M_j)}   \chi_i^s  \right\rbrace ,
    \end{split}
    \label{Kernel_BC}
  \end{equation}
where the label $k$ refers to the intermediate baryon involved in the process. Here, $u$ stands for the Mandelstam variable defined by $u=({p_i}^\mu - {q_j}^\mu )^2$, with ${q_j}^\mu$ being the four-momentum of the final meson and ${p_i}^\mu$ of the initial baryon. The coefficients $C^{(\text{Born})}_{\bar{x}y,z}$, which depend on the axial vector constants $D$ and $F$, can be found in Apendix A of Ref.~\cite{Borasoy:2005ie}.

Taking into account only the tree-level contributions at NLO (diagram (iv) in Fig.~\ref{diagrams}), one obtains:
\begin{equation}
  \begin{split}
    V^{NLO}_{ij}  &= \frac{ N_iN_j}{f^2} \left[ D_{ij}-2L_{ij}{q_j}^\mu {q_i}_\mu +\frac{1}{2M_N^2} g_{ij} ({p_i}^\mu {q_j}_\mu {p_i}^\nu {q_i}_\nu +{p_j}^\mu {q_j}_\mu {p_j}^\nu {q_i}_\nu) \right]  \left(\vphantom{\frac{1}{(E_i+M_i)(E_j+M_j)}}\chi_j^{\dagger s'} \chi_i^s  \right. \\
      &-  \left. \chi_j^{\dagger s'} \frac{\vec{q}_j \cdot \vec{q}_i  + i (\vec{q}_j \times \vec{q}_i ) \cdot \vec{\sigma}}{(E_i+M_i)(E_j+M_j)}   \chi_i^s \right) + \frac{ N_iN_j}{f^2} h_{ij}\left[-\left(\frac{{q_j}_0 {q_i}^2 }{E_i+M_i}+\frac{{q_i}_0 {q_j}^2 }{E_j+M_j}  \right.  \right. \\
    &+ \left.  \left. \frac{{q_j}^2 {q_i}^2 }{(E_i+M_i)(E_j+M_j)} + \frac{(\vec{q}_j \cdot \vec{q}_i)^2}{(E_i+M_i)(E_j+M_j)} \right) \chi_j^{\dagger s'} \chi_i^s  \right.\\
    &+ \left( \frac{{q_i}_0 }{E_i+M_i}+\frac{{q_j}_0}{E_j+M_j} \right)    \chi_j^{\dagger s'} \vec{q}_j \cdot \vec{q}_i \chi_i^s +   \left(  \frac{{q_i}_0 }{E_i+M_i}+\frac{{q_j}_0}{E_j+M_j}    \right. \\
    &+ \left. \left. \frac{ \vec{q}_j \cdot \vec{q}_i}{(E_i+M_i)(E_j+M_j)}-1 \right)  i \chi_j^{\dagger s'} (\vec{q}_j \times \vec{q}_i ) \cdot \vec{\sigma}  \chi_i^s  \right],
  \end{split}
  \label{Kernel_NLO}
\end{equation}
where the $D_{ij}$ and $L_{ij}$ coefficients are constructed from combinations of the NLO parameters $b_0$, $b_D$, $b_F$, $d_1$, $d_2$, $d_3$ and $d_4$ and are given in Table VIII of Ref.~\cite{Feijoo:2015yja}, while the $g_{ij}$ and $h_{ij}$ matrices are listed in Appendix~\ref{ghcoeffs}.

One should bear in mind that the direct incorporation of the amplitudes given by equations~\eqref{Kernel_WT}--\eqref{Kernel_NLO} into Eq.~(\ref{T_algebraic}) is simply not
possible since
such interaction kernels are composed of a mixture of contributions with different angular momenta. Thus,
it is convenient to express the $T$-matrix in terms of the spin-nonflip and spin-flip parts.
\begin{equation}
 T_{ij}(s,s')=\chi_j^{\dagger s'}[f(\sqrt{s},\theta)-i(\vec{\sigma}\cdot \hat{n})g(\sqrt{s},\theta)]\chi_i^s  ,
 \label{eq11}
\end{equation}
where $\theta$ is the CM angle between the inital and final meson momenta and
\(\hat{n}=\vec{q}_j {\times} \vec{q}_i / \lvert\vec{q}_j {\times} \vec{q}_i\rvert\)
is the normal vector to the scattering plane.
The functions $f(\sqrt{s},\theta)$ and $g(\sqrt{s},\theta)$ can be expanded in Legendre polynomials as:
\begin{equation}
  \begin{split}
    f(\sqrt{s},\theta) &=  \sum_{l=0}^\infty f_l (\sqrt{s}) \,  P_l(\cos\theta), \\
    g(\sqrt{s},\theta) &=  \sum_{l=1}^\infty  g_l (\sqrt{s}) \,  \sin\theta \,  \frac{\mathrm{d}P_l(\cos\theta)}{\mathrm{d}\cos\theta},
  \end{split}
  \label{eq12}
\end{equation}
with $f_l$ and $g_l$ being the projections of the sum of all the above defined kernels onto $P_l(\cos\theta)$ and $ \sin\theta \,  \frac{{\rm d}P_l(\cos\theta)}{{\rm d}\cos\theta}$, respectively.

The amplitudes can be redefined as:
\begin{equation}
  \begin{split}
    f_{l+}(\sqrt{s}) &=   \frac{1}{2l+1} \left( f_l (\sqrt{s}) + l \, g_l (\sqrt{s}) \right) \text{for~} J =l+\frac{1}{2}, \\
    f_{l-}(\sqrt{s}) &=    \frac{1}{2l+1} \left( f_l (\sqrt{s}) - (l+1)\,  g_l (\sqrt{s}) \right) \text{for~} J =l-\frac{1}{2},
  \end{split}
  \label{tree_ampl}
\end{equation}
to ensure that the quantum numbers of spin ($\frac{1}{2}$), orbital angular momentum $(l)$, and total angular momentum $(J)$ are preserved in the unitarization procedure implemented by the BS equations. Following the notation of Ref.~\cite{Jido:2002zk}, each unitarized $J$-scattering amplitude should be calculated by a new version of Eq.~(\ref{T_algebraic}), which in matrix form reads:
\begin{equation}
  f_{l\pm}  =   \left[  1 - f^{\rm tree}_{l\pm}G \right]^{-1}  f^{\rm tree}_{l\pm},
\end{equation}
where the amplitudes $f^{\rm tree}_{l\pm}$ are obtained from equations~(\ref{eq11})--(\ref{tree_ampl}) but employing, for the amplitude $T_{ij}(s,s')$, the sum of the interaction kernels shown in equations (\ref{Kernel_WT})--(\ref{Kernel_NLO}).

The aim of this work is the study of the effects caused by the inclusion of higher partial waves on the physical observables in this sector, particularly on the $K^- p$ cross sections.
The most general expression of the differential cross sections for a given $\phi_i B_i^s \to \phi_j B_j^{s'}$ process is
\begin{equation}
  \frac{d\sigma_{ij}}{d\Omega}=\frac{M_i \, M_j \, q_j}{16 \,  \pi^2 \, s \,  q_i} \,  S_{ij},
\end{equation}
with
\begin{equation*}
  S_{ij}= \frac{1}{2s+1}  \sum_{s,s'}\, \lvert T_{ij}(s,s')\rvert^2 = \lvert f(w,\theta)\rvert^2 +  \lvert g(w,\theta)\rvert^2,
\end{equation*}
where the first factor averages over the initial baryon spin projections, giving $1/2$ for this particular case, and where we have also summed over all possible final baryon spin projections.

Since the scope of the present study does not go beyond d-wave contributions, focusing mostly on p-wave effects, the
expression for the total cross section incorporating such partial waves can be written as:
\begin{equation}
\label{tot_xsect}
 \sigma_{ij}=\frac{M_i \, M_j \, q_j}{4 \,  \pi \, s \,  q_i} \, \left[  \lvert f_{0}\rvert^2+ 2 \lvert f_{1+}\rvert^2 + \lvert f_{1-}\rvert^2 +
 3 \lvert f_{2+} \rvert^2 + 2\lvert f_{2-}\rvert^2 \right] .
\end{equation}

\section{Models and Data Treatment}
\label{sec:modelsData}

The availability of a large collection of experimental data makes the $S=-1$ sector an interesting benchmark to test meson-baryon EFTs, not only for checking their predictive power but also to extract information about the LECs,
especially those beyond LO.
Constructing a model with higher partial waves capable to describe the phenomenology of the $S=-1$ meson--baryon interaction provides, indeed, a tool to explore other sectors lacking experimental data, like $S=-2$, $S=-3$, and even to check the validity of SU(3) symmetry upon comparing the resulting parameters to those of existing chiral models that describe the meson--baryon interaction in the $S=0$ sector.

\subsection{Models}

In this section, we introduce the models derived for this study
and discuss the fitting procedure. Since the present study is the natural extension of the one presented in~\cite{Feijoo:2018den}, we include a description of the old model as a reminder and to provide a better comprehension of the p-wave effects.

\begin{itemize}[leftmargin=*,labelsep=5.8mm]

\item {\bf s-wave (old)}: This first model corresponds to the fit called \textbf{WT+Born+NLO} carried out in~\cite{Feijoo:2018den}. It was constructed by adding the interaction kernels derived from the Lagrangian up to NLO and neglecting the \(h\)-{} and \(g\)-terms. We limited this model to the s-wave contribution.

\item {\bf s-wave}: The second model improves upon the first one by incorporating the novel \(h\)-{} and \(g\)-terms that come from the NLO interaction kernel and, as in the first model, only the s-wave contribution is taken into account.

\item {\bf s+p-waves}: This model employs the same Lagrangian as the {\bf s-wave} fit, but it also incorporates the p-wave contributions.

\end{itemize}

\subsection{Data Treatment and Fitting Procedure}

The parameters in our models originate from the chiral Lagrangian as well as from the employed regularization scheme. One finds 17 LECs in accordance with equations~\eqref{Kernel_WT}--\eqref{Kernel_NLO}, namely the meson decay constant $f$, the axial vector couplings $D$ and $F$, and the NLO coefficients $b_0$, $b_D$, $b_F$, $d_1$, $d_2$, $d_3$, $d_4$, $g_1$, $g_2$, $g_4$, $h_1$, $h_2$, $h_3$, and $h_4$. It is a well-known fact that, in the UChPT models, the $f$ parameter effectively takes larger values than the experimental one. Literature offers plenty of miscellaneous choices but, in our particular case, in order to
allow for a sort of
average over all mesons involved in the various coupled channels, this parameter is constrained to vary from $f=f_{\pi}$ to $f=1.3 f_{\pi}$. The axial vector couplings $D$ and $F$
are allowed to vary within 15\% of their canonical values in order to accommodate the spread of values found in the literature~\cite{Ramos:2016odk}. The NLO LECs are treated as completely free parameters in the fit. Additionally, the dimensional regularization introduces six subtraction constants, $a_{\pi \Sigma}$, $a_{\bar{K} N}$, $a_{\pi \Lambda}$, $a_{\eta \Sigma}$, $a_{\eta \Lambda}$, and $a_{K \Xi}$ that are also parameters of the fit.
However, these parameters are restricted to their natural-sized values, which should, as discussed in Ref.~\cite{Feijoo:2018den}, lie in the range $[-10, 10]\times 10^{-3}$ for
$\mu = 1$~GeV.

\begin{table}
\caption{Number of experimental points used in our fits, which are extracted from~\cite{exp_1,exp_2, exp_3, exp_4, exp_5, exp_6, exp_7, exp_8, exp_9, exp_10, exp_11,br_1,br_2,SIDD,Starostin,Baxter,Jones,Berthon,Jones:1974si}, distributed per observable.}
\label{tab_exp_points}
\centering
\begin{tabular}{lclc}
\hline \\[-2.5mm]
 {Observable}\, & \, {Points} \, & \, {Observable}\, & \, {Points} \\
\hline \\[-2.5mm]
$\sigma_{K^-p \to K^-p}$ & 23 & $\sigma_{K^-p \to \bar{K}^0n}$ & 9 \\
$\sigma_{K^-p \to \pi^0\Lambda}$ & 3 & $\sigma_{K^-p \to \pi^0\Sigma^0}$ & 3 \\
$\sigma_{K^-p \to \pi^-\Sigma^+}$ & 20 & $\sigma_{K^-p \to \pi^+\Sigma^-}$ & 28 \\
$\sigma_{K^-p \to \eta\Sigma^0}$ & 9 & $\sigma_{K^-p \to \eta\Lambda}$ & 49 \\
$\sigma_{K^-p \to K^+\Xi^-}$ & 46 & $\sigma_{K^-p \to K^0\Xi^0}$ & 29 \\
$\gamma$ & 1 & $\Delta E_{1s}$ & 1 \\
$R_n$ & 1 & $\Gamma_{1s}$ & 1 \\
$R_c$ & 1 &  & \\
\hline
\end{tabular}
\end{table}

Table \ref{tab_exp_points} displays the distribution per observable of the 224 experimental points employed in the current fitting procedure (the same as in~\cite{Feijoo:2018den}), most of them coming from the total cross section for $K^-p$ scattering into different final channels~\cite{exp_1,exp_2, exp_3, exp_4, exp_5, exp_6, exp_7, exp_8, exp_9, exp_10, exp_11,Starostin,Baxter,Jones,Berthon,Jones:1974si}. Apart from this, the measured branching ratios of cross section yields~\cite{br_1,br_2} are also fitted. Such yields can be defined from the elastic and inelastic $K^-p$ cross sections evaluated at threshold:
\begin{equation}
  \begin{split}
    \gamma  &=  \frac{\Gamma(K^- p \rightarrow \pi^+ \Sigma^-)}{\Gamma(K^- p \rightarrow \pi^- \Sigma^+)}= 2.36 \pm 0.04 ,\\
    R_c &= \frac{\Gamma(K^- p \rightarrow \pi^+ \Sigma^-,\pi^- \Sigma^+ )}{\Gamma(K^- p \rightarrow \text{inelastic channels})}=0.189 \pm 0.015 , \\
    R_n &= \frac{\Gamma(K^- p \rightarrow \pi^0 \Lambda)}{\Gamma(K^- p \rightarrow \text{neutral states})} = 0.664 \pm 0.011 .
  \end{split}
\label{branch_ratios}
\end{equation}

The
precise measurement of the energy level shift and width of the atomic $1s$ state in kaonic hydrogen by the SIDDHARTA Collaboration~\cite{SIDD} implied a substantial change in constraining the theoretical models. We incorporate this input in our fits via the $K^- p$ scattering length, which is obtained from the $K^- p$ scattering amplitude at threshold as:
\begin{equation}
  a_{\scriptscriptstyle K^- p}=-\frac{1}{4\pi}\frac{M_p}{M_p+M_{\bar K}}T_{K^- p \to K^- p}.
  \label{scat_lenght}
\end{equation}
The scattering length is connected to the shift and width of kaonic hydrogen by means of the second order corrected Deser-type formula~\cite{Meissner:2004jr}:
\begin{equation}
\Delta E-i\frac{\Gamma}{2}=-2\alpha^3\mu_{r}^{2} a_{\scriptscriptstyle K^- p} \Big[ 1+2 a_{\scriptscriptstyle K^- p}\,\alpha\,\mu_r\, (1-\ln\alpha) \Big],
\label{ener_shift_width}
 \end{equation}
where $\alpha$ is the fine-structure constant and $\mu_r$ the reduced mass of the $K^-p$ system.

The minimizing criteria taken for the fitting procedure is based on the $\chi^2$ per degree of freedom ($\chi^2_{\rm d.o.f.}$). We avoid the use of the the standard definition of $\chi^2_{\rm d.o.f.}$ since it favors observables with a larger number of experimental points over those with a smaller amount. To get around this misleading effect, we adopt the method used in~\cite{GarciaRecio:2002td,GO,IHW,Mai:2014xna,Feijoo:2015yja, Ramos:2016odk, Feijoo:2018den}, consisting of the assignment of equal weights to the different measurements by defining a renormalized $\chi^2_{\rm d.o.f.}$ as:
\begin{equation}
  \chi^{2}_{\rm d.o.f} = \frac{\sum_{k=1}^K n_k}{\sum_{k=1}^K n_k -p} \frac{1}{K} \sum_{k=1}^K \frac{1}{n_k} \chi^{2}_{k},
  \label{Chi^2_dof}
\end{equation}
with
\begin{equation*}
 \chi^{2}_{k}=\sum_{i=1}^{n_k}\frac{\left( y_{i,k}^{\rm th}- y_{i,k}^{\rm exp}\right)^2}{\sigma_{i,k}^{2}}.
\end{equation*}
Here, $y_{i,k}^{\rm exp}$, $y_{i,k}^{\rm th}$, and $\sigma_{i,k}$ are
the experimental value, the theoretical prediction, and the
experimental uncertainty of the $i^{\text{th}}$ point of the $k^{\text{th}}$
observable, respectively. Each observable has a total of $n_{k}$
points; $K$ is the total number of observables, and $p$ denotes the
number of free fit parameters. In Eq.~(\ref{Chi^2_dof}), the
renormalization is incorporated by averaging the $\chi^2$ per degree
of freedom over the different experiments.


\section{Results and Discussion}
\label{sec:results}
As mentioned in the introduction, our previous works~\cite{Feijoo:2015yja, Ramos:2016odk, Feijoo:2018den} studied the influence of the Born and NLO contributions, especially on the inelastic processes that open up at higher energies, namely the $K^- p\to \eta\Lambda, \eta\Sigma^0, K^+\Xi^-, K^0\Xi^0$ reactions. In the present work we include new ingredients, such as the \(h\)-{} and \(g\)-terms of the NLO Lagrangian and the p-wave contributions, which are expected to play a relevant role at these energies. The effects of these
new contributions
on the model parameters are displayed in Table~\ref{tab:outputs_fits}, where the
parameters of the
\textbf{s-wave (old)}
model%
, obtained in Ref.~\cite{Feijoo:2018den}, are also included
for comparison.

\begin{table}[h]
  \caption{
    Values of the parameters and the corresponding $\chi^2_{\rm d.o.f.}$, defined in Eq.~(\ref{Chi^2_dof}), for the different models described in the text. The subtraction constants are taken at a regularization scale $\mu=1$~GeV. The error bars in the parameters of the \textbf{s-wave (old)} are determined as explained in~\cite{Feijoo:2018den}, while those of \textbf{s+p-waves} and \textbf{s-wave} are directly provided by the MINUIT~\cite{James:1975dr} minimization procedure.}
  \label{tab:outputs_fits}
\centering
\begin{tabular}{lccc}
\hline \\[-2.5mm]
                                                 & \textbf{s+p-waves}                    &  \textbf{s-wave}                 & \textbf{s-wave (old)}                                          \\
\hline \\[-2.5mm]
$a_{\bar{K}N} \ (10^{-3})$       & $1.322 \pm 0.643$       & $ 2.105 \pm 0.378$    & $ 1.268^{+0.096}_{-0.096}$    \\
$a_{\pi\Lambda}\ (10^{-3})$    & $ 10.000 \pm 19.962$     & $ 9.999 \pm 18.190 $   & $ -6.114^{+0.045}_{-0.055}$    \\
$a_{\pi\Sigma}\ (10^{-3})$       & $ 1.247 \pm 1.913$       &  $ 3.413 \pm 1.609$    & $ 0.684^{+0.429}_{-0.572}$    \\
$a_{\eta\Lambda}\ (10^{-3})$  & $ -3.682\pm 6.541$     &  $ -4.585 \pm 1.322 $  & $ -0.666^{+0.080}_{-0.140}$     \\
$a_{\eta\Sigma}\ (10^{-3})$     & $ 5.528 \pm 2.571$       &  $3.017 \pm 0.027$    & $ 8.004^{+2.282}_{-0.978}$    \\
$a_{K\Xi}\ (10^{-3})$                & $ -2.077\pm 0.931$       &  $0.997 \pm 0.038$        & $ -2.508^{+0.396}_{-0.297}$   \\
$f/f_{\pi}$                                & $1.110 \pm 0.068$    &  $ 1.042\pm 0.003$       & $ 1.196^{+0.013}_{-0.007}$         \\
$b_0 \ (GeV^{-1}) $                  & $ 0.394 \pm 0.087$       &  $-0.079 \pm 0.005$      & $ 0.129^{+0.032}_{-0.032}$      \\
$b_D \ (GeV^{-1}) $                 & $ 0.206 \pm 0.064$        &  $ 0.112\pm 0.008$      & $ 0.120^{+0.010}_{-0.009}$      \\
$b_F \ (GeV^{-1}) $                 & $ 0.303 \pm 0.058$       &   $0.117 \pm 0.011$      & $ 0.209^{+0.022}_{-0.026}$      \\
$d_1 \ (GeV^{-1}) $                & $ 0.246 \pm 0.077$      &   $-0.848 \pm 0.039$     & $ 0.151^{+0.021}_{-0.027}$       \\
$d_2 \ (GeV^{-1}) $                & $ 0.120 \pm 0.057$    &   $ 0.634\pm 0.036$     & $ 0.126^{+0.012}_{-0.009}$        \\
$d_3 \ (GeV^{-1}) $                & $ 0.270\pm 0.082$        &   $0.463 \pm 0.047$   & $ 0.299^{+0.020}_{-0.024}$       \\
$d_4 \ (GeV^{-1}) $                & $ 0.723 \pm 0.085$     &   $ -0.678 \pm 0.032$     & $ 0.249^{+0.027}_{-0.033}$       \\
$g_1 \ (GeV^{-1})$                 & $ 0.105 \pm 0.114$         &  $-1.211 \pm 0.049$      & \multicolumn{1}{c}{-}                \\
$g_2 \ (GeV^{-1})$                & $ -0.024 \pm 0.056$            &   $ 0.764 \pm 0.041 $      & \multicolumn{1}{c}{-}            \\
$g_4 \ (GeV^{-1})$                & $ 0.301\pm 0.097 $                   &  $ -1.030 \pm 0.036$    & \multicolumn{1}{c}{-}           \\
$h_1 \ (GeV^{-1})$                & $ 0.540 \pm 1.070$              &  $ -0.533 \pm 0.373$         & \multicolumn{1}{c}{-}            \\
$h_2 \ (GeV^{-1})$                 & $ 0.387\pm 0.483$                &  $-1.979 \pm 0.229$    & \multicolumn{1}{c}{-}              \\
$h_3 \ (GeV^{-1})$                 & $ 0.472 \pm 0.821$              &  $ 7.452 \pm 0.159$     & \multicolumn{1}{c}{-}                \\
$h_4 \ (GeV^{-1})$                 & $ -0.291\pm 0.832$               &  $ -2.547 \pm 0.319 $       & \multicolumn{1}{c}{-}             \\
$D$                                        & $ 0.701 \pm 0.102$      &   $0.899 \pm 0.004$      & $ 0.700^{+0.064}_{-0.144}$           \\
$F$                                        & $ 0.510 \pm 0.056$    &     $ 0.510\pm 0.017$       & $ 0.510^{+0.060}_{-0.050}$          \\

\hline \\[-2.5mm]
$\chi^2_{\rm d.o.f.}$              & $0.86$                             &  $0.77$                      &  $1.14$                         \\[2pt]
\hline
\end{tabular}
\end{table}

The parameters of the new \textbf{s-wave} model are displayed in the second column of Table~\ref{tab:outputs_fits}. Although this fit has the lowest $\chi^2_{\text{d.o.f.}}$ of $0.77$, we observe
an overall discrepancy in the values of the parameters and in their accuracy compared to those obtained in our previous \textbf{s-wave (old)} study. It is important to stress that
the error bars can not be directly compared because of the different methods employed to calculate them, as mentioned in the caption of Table~\ref{tab:outputs_fits}. Despite this fact, some qualitative comparison can be made, as we discuss below.

Although we get natural-sized values for all the subtraction constants obtained with the two s-wave models,
the parameter sets are very different.
Moreover, the \textbf{s-wave}
model provides larger values of the $d_i$ parameters compared to those of the \textbf{s-wave (old)}
model, quite high values of the new $g_i$'s and $h_i$'s, as well as a small value of $f/f_\pi$, almost at the lower edge, together with values of the axial vector constants at their upper edges. A plausible explanation is that the \(g\)-{} and \(h\)-terms of the new model can play a role in accommodating the data at higher energies, where these terms are more important, and the fitting procedure forces them to be sizable at the expense of producing extreme values for the other parameters.

It is therefore expected that the incorporation of the p-waves in the fit, acting also more importantly at higher energies~\cite{CaroRamon:1999jf}, might bring the parameters of the \(g\)-{} and \(h\)-terms within a more natural range. This is indeed what happens for the \textbf{s+p-waves} model, whose parameters are compiled in the first column of Table~\ref{tab:outputs_fits}. The parameters $b_i$'s and $d_i$'s become also more moderate in size, except for the value of $d_4$ which shows big differences among the different models. We note that the volatility of this NLO LEC was already discussed in~\cite{Feijoo:2018den} and it was tied to the fact that this parameter only plays a role in the elastic transitions
of the sector. Finally, we observe that the fit of the new \textbf{s+p-waves} model produces a value of $f/f_\pi$ very close to that of the old model, together with exactly the same values for $D$ and $F$.

\begin{table*}[!ht]
  \caption{Threshold observables obtained from our fits. Experimental data is taken from~\cite{br_1,br_2,SIDD}. }
  \label{tab:thresh}
\centering
\begin{tabular}{lcccccc}
\hline \\[-2.5mm]
                             & {$\gamma$} & {$R_n$} & {$R_c$}&  {$a_p(K^-p \rightarrow K^- p)$}  & {$\Delta E_{1s}$}     & {$\Gamma_{1s}$} \\
\hline \\[-2.5mm]
\textbf{s+p-waves}  & 2.36   &  0.188    & 0.662    & $-0.70+{\rm i\,}0.81$                   & 297                         & 532  \\%
\textbf{s-wave}   &  2.40  & 0.179  &    0.665   & $- 0.64+{\rm i\,}0.83$                &   280                       & 560   \\%
\textbf{s-wave (old)}             & $2.36^{+0.03}_{-0.03}$ & $0.188^{+0.010}_{-0.011}$ & $0.659^{+0.005}_{-0.002}$   & $-0.65^{+0.02}_{-0.08}+{\rm i\,}0.88^{+0.02}_{-0.05}$                   & $288^{+23}_{-8}$                         & $588^{+9}_{-40}$\\%

\hline \\[-2.5mm]
Exp.    & $2.36{\pm} 0.04$ &	$0.189{\pm} 0.015$ & $0.664{\pm} 0.011$        & $ (-0.66{\pm}0.07) {+} {\rm i\,}(0.81{\pm}0.15)$ &	$283{\pm}36$ & $541{\pm}92$ \\
\hline
\end{tabular}
\end{table*}

The results of the threshold observables for the previous models are collected in Table~\ref{tab:thresh}, where they can be compared to the corresponding experimental values also included there. There is no substantial change in the reproduction of the experimental values when comparing the new models to the {\bf s-wave (old)} one.

\begin{figure}[ht!]
\centering
 \includegraphics[width=6.2 in]{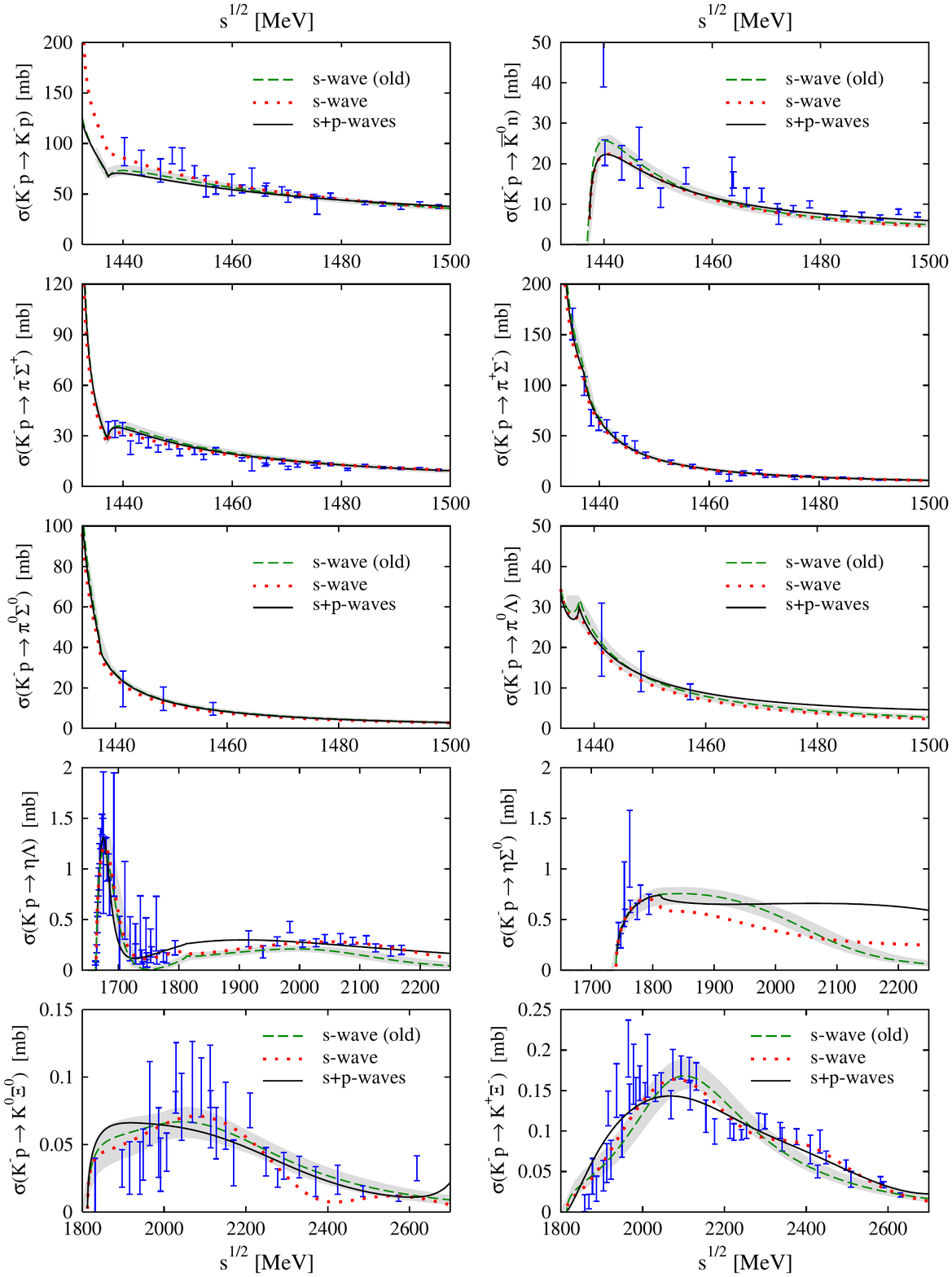}
\caption{Total cross sections of the $K^- p\to K^- p, \bar{K}^0n, \pi^- \Sigma^+, \pi^+\Sigma^-, \pi^0 \Sigma^0, \pi^0\Lambda, \eta\Lambda, \eta\Sigma^0, K^+\Xi^-, K^0\Xi^0$ reactions obtained for the {\bf s+p-waves} (solid black line), {\bf s-wave (old)} (dashed green line), with the corresponding estimation of the error bands (grey), and for {\bf s-wave} (dotted red line) models (dashed line). Experimental data has been taken from~\cite{exp_1,exp_2, exp_3, exp_4, exp_5, exp_6, exp_7, exp_8, exp_9, exp_10, exp_11,Starostin,Baxter,Jones,Berthon,Jones:1974si}. See the text for a detailed description of the models.}
  \label{fig:models_comp}
\end{figure}

In Fig.~\ref{fig:models_comp}, the calculated total cross sections of $K^-p$ scattering to all channels of the $S=-1$ sector are compared to experimental data. As can be seen from the first three panel rows, the agreement of all models in reproducing the threshold observables is also reflected on the total cross sections of the classical processes, namely $K^- p\to K^- p$, $\bar{K}^0n$, $\pi^- \Sigma^+$, $\pi^+\Sigma^-$, $\pi^0 \Sigma^0$, and $\pi^0\Lambda$, which are studied at energies close to the $K^- p$ threshold.
It is interesting to comment on the little enhancement of strength seen for the {\bf s+p-waves} model at CM energies around $1500$~MeV in the $K^- p \to \pi^0\Lambda$ cross section. In this case, since the corresponding threshold is located several tens of MeV below the other channels, the momentum of the outgoing particles can reach higher values thereby providing a more relevant role to the p-wave contribution.

In contrast to what we see in the conventional channels, larger differences among the models are observed in the cross sections of the $K^- p \to \eta\Lambda, \eta\Sigma^0, K^+\Xi^-, K^0\Xi^0$ processes, shown in the two bottom panel rows of Fig.~\ref{fig:models_comp}. In the case of the $\eta$ channels, one can clearly see that all models describe qualitatively well the data, exhibiting minor discrepancies at energies below $1800$~MeV. The differences among the models become appreciable at higher energies.
From the study in~\cite{GO,Feijoo:2018den}, we know that the $\eta\Lambda$ and $\eta\Sigma^0$ reactions are sensitive to the NLO (and Born) terms up to the extent that one cannot reproduce properly the experimental data in the case of $\eta\Lambda$ channel at this energy range unless such terms are taken into account. Consequently, the disagreement observed between the {\bf s-wave (old)} and {\bf s-wave} models can be directly attributed to effects induced by the incorporation of the new \(g\)-{} and \(h\)-terms.
The additional p-wave contribution in the {\bf s+p-waves} model, which gains significance as the momenta of particles involved increase, seems to provide more strength to the $K^- p \to \eta\Lambda$ cross section around $1900$~MeV and to the $K^- p \to \eta\Sigma^0$ one from $2000$~MeV on.

In the case of the $K\Xi$ channels, there are appreciable differences between the models in the whole energy range. In particular, the {\bf s+p-waves}
model
for the $K^0\Xi^0$ cross section slightly
overestimates
the experimental data below $2000$~MeV while remaining somewhat below the experimental structure that appears at $2100$~MeV. On the contrary, this model offers the best description for the experimental points located at higher energies for both the $K^0\Xi^0$ and $K^+\Xi^-$ total cross sections. This is in opposition to the {\bf s-wave} and {\bf s-wave (old)}
models
that, in such a region, accommodate the experimental points of the $K^0\Xi^0$ and $K^+\Xi^-$ cross sections, respectively, at the edges of their errors bars. In general, the {\bf s+p-waves} model is smoothly averaging over the whole set of points and the s-wave models are able to provide a more marked structure around 2100~MeV.

\begin{figure}[ht!]
\centering
 \includegraphics[width=6.2 in]{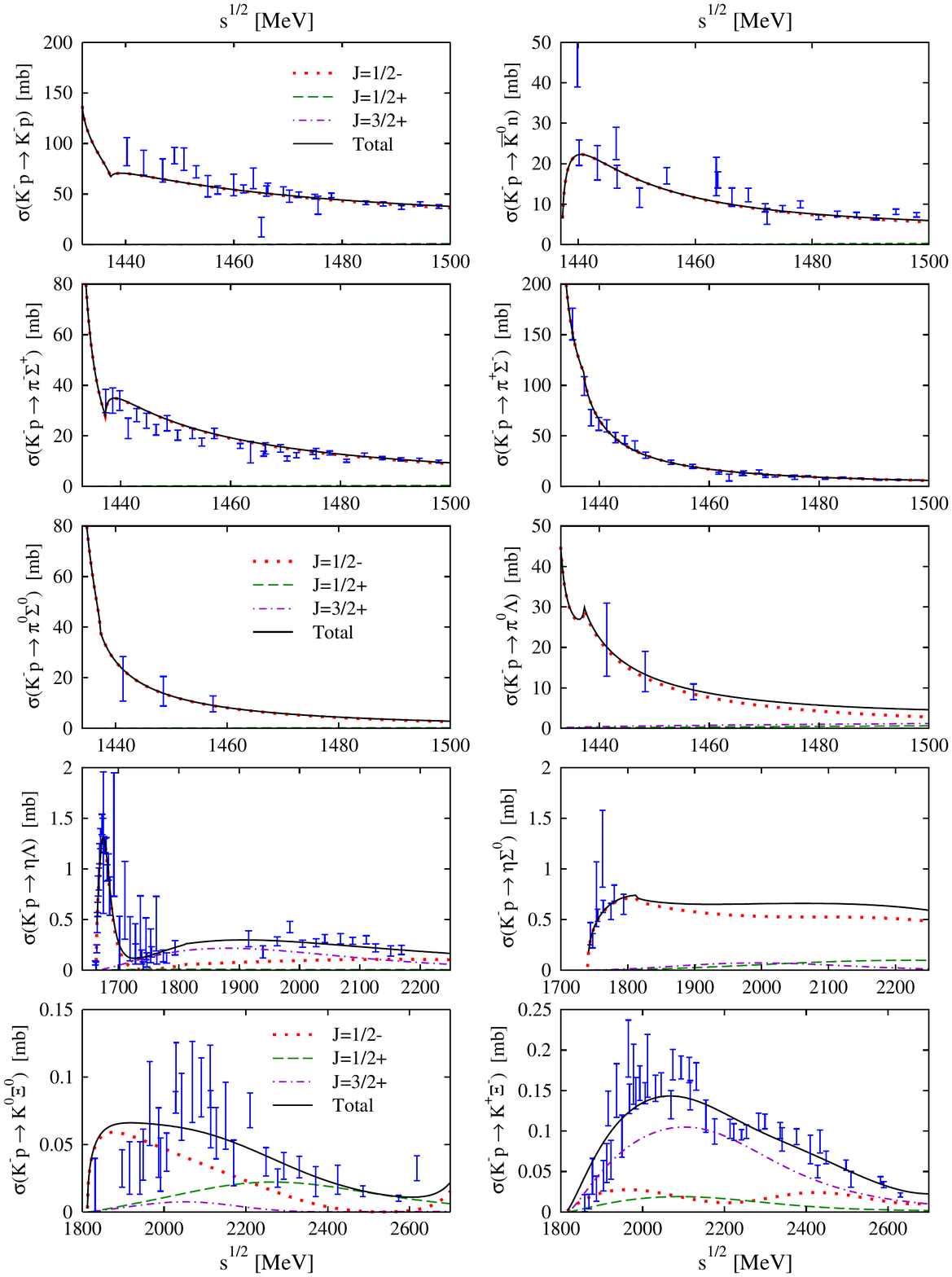}
\caption{Total cross sections of the $K^- p\to K^- p, \bar{K}^0n, \pi^- \Sigma^+, \pi^+\Sigma^-, \pi^0 \Sigma^0, \pi^0\Lambda, \eta\Lambda, \eta\Sigma^0, K^+\Xi^-, K^0\Xi^0$ reactions obtained for {\bf s+p-waves} (solid black line), and the corresponding contributions for $J^P={\frac{1}{2}}^-$ (dotted red line), $J^P={\frac{1}{2}}^+$ (dashed green line) and $J^P={\frac{3}{2}}^+$ (dash-dotted violet line).}
  \label{fig:xsect_P_contr}
\end{figure}

For a deeper understanding
of the role of the p-waves, in Fig.~\ref{fig:xsect_P_contr} we present the individual partial-wave contributions to the $K^-p$ processes included in the {\bf s+p-waves} model, namely the $J^P={\frac{1}{2}}^-$ (s-wave), ${\frac{1}{2}}^+$ and ${\frac{3}{2}}^+$ (p-wave) channels. Turning first to the classical processes (top six panels), we notice that p-wave effects are only noticeable in the $\pi^0\Lambda$ production process at the upper edge of the energy range, as discussed in the previous paragraph. The other cross sections are essentially built from s-wave contributions at the energies explored here.

The situation becomes much more interesting for those
reaction channels
that open up at higher energies. Examining the $K^- p \to \eta\Lambda$ cross section, one can observe that the s-wave $J^P={\frac{1}{2}}^-$ contribution (dotted red line) is dominant just above threshold due to the $\Lambda(1670)$ resonance, generated dynamically in this partial wave. Then, as the energy increases, its role is moderately losing relevance against the $J^P={\frac{3}{2}}^+$ component (dash-dotted violet line) which is the main contribution in the region ranging from $1750$ to $2100$~MeV. As a matter of fact, the predominant character of the $J^P={\frac{3}{2}}^+$ component is another proof of the need to incorporate NLO terms in the kernel, as it can be seen that the WT and the Born s-channel terms do not contribute to this partial wave.
Moreover, it is interesting to note that the present $J^P={\frac{3}{2}}^+$ contribution may replace the role of the explicit resonance $\Lambda(1890)$ with these quantum numbers that was incorporated, in addition to the s-wave chiral amplitudes, in one of the models studied in~\cite{Feijoo:2018den} producing a similar description of the $\eta\Lambda$ cross section above $1900$~MeV. Concerning the $J^P={\frac{1}{2}}^+$ partial-wave (dashed green line), it is hardly contributing to the cross section which is in contrast to the coupled channel partial wave analysis carried out in Ref.~\cite{Matveev:2019igl}. There, the corresponding cross section contains a contribution coming from the $J^P={\frac{1}{2}}^+$ channel comparable in size to that of the $J^P={\frac{3}{2}}^+$ one, the latter having a similar strength as ours. As a last comment, we emphasize that the information encoded in this process can be very valuable as it acts as an $I=0$ filter, thereby helping to reduce the potential ambiguities in the amplitudes~\cite{Feijoo:2018den, Feijoo:2015cca}.

Next, we focus on another isospin filtering process, namely the $K^- p \to \eta\Sigma^0$ reaction which proceeds in the $I=1$ channel. In this case, the s-wave $J^P={\frac{1}{2}}^-$ contribution is clearly the dominant one over the whole range of energies explored. The p-wave contributions start being
noticeable from $1850$~MeV, representing 20\% of the total cross section at most. However, one should be very
careful
about drawing conclusions because of the lack of experimental data in this channel beyond $1800$~MeV.

The partial-wave decomposition of the $K\Xi$ total cross section provides very valuable information about the relevance of higher partial waves in our model. On the one hand, the $K^0\Xi^0$ production discloses the fundamental role played by the $J^P={\frac{1}{2}}^+$ contribution above $2200$~MeV and how the low-energy regime is dominated by the $J^P={\frac{1}{2}}^-$ wave. The $J^P={\frac{3}{2}}^+$ component provides a nonnegligible strength around 2100~MeV, which is however not enough to reproduce the sizable experimental structure. On the other hand, in the $K^- p \to K^+\Xi^-$ process, the $J^P={\frac{1}{2}}^-$ and $J^P={\frac{1}{2}}^+$ contributions are moderate and roughly constant, being the $J^P={\frac{3}{2}}^+$ component the one that governs the description of the experimental data. A subdominance of the s-wave component was also obtained for the $\pi^+ p \to K^+ \Sigma^+$ process in Ref.~\cite{CaroRamon:1999jf}, where a similar chiral scheme up to NLO for the $S=0$ sector was employed.
The resonant model of Ref.~\cite{Matveev:2019igl} produces $K^0\Xi^0$ and $K^+\Xi^-$ cross section that are dominated by the ${\frac{3}{2}}^+$ contribution with a sizable influence of the ${\frac{5}{2}}^-$ one. The s-wave $J^P={\frac{1}{2}}^-$ component is weak and the ${\frac{1}{2}}^+$ contribution practically nonexistent, in strong opposition to the present results.


\begin{figure}[ht!]
\centering
 \includegraphics[width=6.2 in]{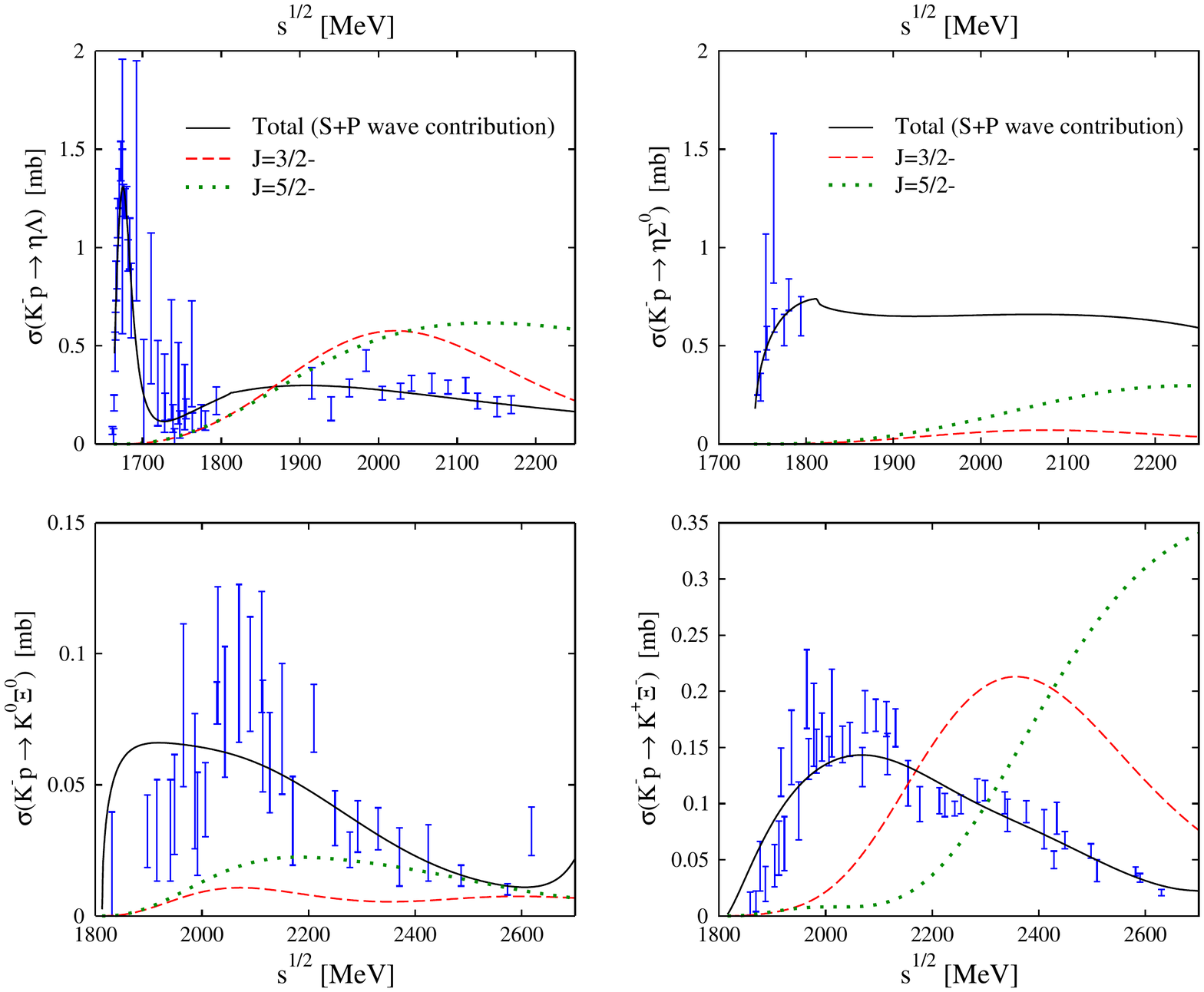}
\caption{Total cross sections of the $K^- p\to \eta\Lambda, \eta\Sigma^0, K^+\Xi^-, K^0\Xi^0$ reactions obtained for {\bf s+p-waves} (solid black line) compared to the contributions for $J^P={\frac{3}{2}}^-$ (dashed red line) and $J^P={\frac{5}{2}}^-$ (dotted green line) one obtains employing the same parametrization (first column in Table~\ref{tab:outputs_fits}). See text for a detailed explanation of the content.}
  \label{fig:xsect_D_contr}
\end{figure}

The incorporation of the d-wave components of the interaction is the natural next step
It can be easily
verified
that these partial waves emerge from the crossed Born and the NLO kernels.
It seems that
a model containing the s-, p-{} and d-wave components simultaneously would provide the most realistic information about the NLO LECs of the Lagrangian (\ref{LagrphiB2}). 
However, in practice, reaching a proper parametrization of the LECs when the d-waves are included is highly non-trivial, given the interdependence of the parameters contributing to the various partial waves at tree level, together with the complications of the coupled-channel scheme. Preliminary attempts have led us to the conclusion that a proper and controlled inclusion of higher partial waves in the fit requires the extension of the data set to include the available total and differential cross sections at higher energies, an enormous task that is beyond the scope of the present work.

However, in order to illustrate the relevance of the d-waves, we take the parameters of the {\bf s+p-waves}
model
and present, in Fig.~\ref{fig:xsect_D_contr}, the $J^P={\frac{3}{2}}^-$ and $J^P={\frac{5}{2}}^-$ contributions to the cross sections of the $K^- p\to \eta\Lambda, \eta\Sigma^0, K^+\Xi^-, K^0\Xi^0$ reactions, which are the ones expected to be more substantially affected by the higher partial waves. As can be noticed from all panels in Fig.~\ref{fig:xsect_D_contr}, the d-wave contributions start to have a non-negligible effect about 100~MeV above the threshold of the corresponding reaction. In the cases of the $\eta\Lambda$ and $K^+\Xi^-$ cross sections these partial waves contribute in a very remarkable way. This demonstrates that the d-wave terms seem to be an essential ingredient to describe the $K^- p$ inelastic cross sections at higher energies and, consequently, a fitting procedure including the effect of the d-wave terms needs to be performed. From a purely phenomenological point of view, it seems that the $J^P={\frac{3}{2}}^-$ and $J^P={\frac{5}{2}}^-$ contributions could indeed provide the additional structures needed to accommodate the experimental data.
But the most important consequence of the inclusion of such contributions is the possibility to get more reliable values for the NLO LECs. This prospect stems from the fact that the d-waves receive contributions mostly from the NLO terms of the Lagrangian, whereas the d-wave contributions of the LO terms come exclusively from the crossed Born diagram.

\begin{table*}[t]
  \caption{
    Comparison of the pole positions between the models: {\bf s+p-waves}, {\bf s-wave (old)} and {\bf s-wave} (in MeV) and the corresponding modulus of the couplings $\lvert g_i\rvert$ found in $I=0$ and $I=1$ channels.}
  \label{tab:spectroscopy}
\centering
\begin{tabular}{c|cc|c|c|c}
\hline
\hline
\multicolumn{6}{c}{ {\bf s+p-waves} } \\
\hline
  {\bf $J^P=\frac{1}{2}^-$}  &\multicolumn{3}{|c}{{\bf$(I,S)=(0,-1)$}} &  \multicolumn{2}{|c}{{\bf$(I,S)=(1,-1)$}}  \\
  [0.5mm]
\hline
   &    \multicolumn{2}{c|}{$\Lambda(1405)$}    &  $\Lambda(1670)$ &  & $\Sigma^*$ \\
  &     &     &  &   &   \\
  [-4.5mm]
$M\;\rm[MeV]$             &  $1399.71 $    &  $1423.30$   & $1674.05$ & $M\;\rm[MeV]$  & $1590.97$  \\
$\Gamma\;\rm[MeV]$   &  $118.50$    &  $58.02$   & $31.08$ & $\Gamma\;\rm[MeV]$ & $ 480.14$ \\
\hline
                                &   $\lvert g_i\rvert$   &   $\lvert g_i\rvert$   &  $\lvert g_i\rvert$  &   & $\lvert g_i\rvert$        \\
$\pi \Sigma$            &  $3.45$     &  $2.57$   &   $0.37$  & $\pi \Lambda$   &  $1.24$  \\
$\bar{K}N$              &  $3.19$     &  $3.70$   &   $0.40$   & $\pi \Sigma$     &  $ 1.36$   \\
$\eta \Lambda$       &  $0.66$     &  $0.85$   &   $1.33$   & $\bar{K}N$       &   $1.79$   \\
$K \Xi$                    &  $0.72$    &  $0.52$    &  $3.65$    &  $\eta \Sigma$  & $0.37$   \\
                               &                 &                 &                 & $K \Xi$              &  $0.57$    \\
\hline
\hline
\multicolumn{6}{c}{ {\bf s-wave} } \\
\hline
  {\bf $J^P=\frac{1}{2}^-$}  &\multicolumn{3}{|c}{{\bf$(I,S)=(0,-1)$}} &  \multicolumn{2}{|c}{{\bf$(I,S)=(1,-1)$}}  \\
  [0.5mm]
\hline
   &    \multicolumn{2}{c|}{$\Lambda(1405)$}    &  $\Lambda(1670)$ &  & $\Sigma^*$ \\
  &     &     &  &   &   \\
  [-4.5mm]
$M\;\rm[MeV]$             &  $1364.13$    &  $1419.54$   & $1679.16$ & $M\;\rm[MeV]$  & $-$  \\
$\Gamma\;\rm[MeV]$   &  $190.58$    &  $ 39.14$   & $62.36$ & $\Gamma\;\rm[MeV]$ & $-$ \\
\hline
                                &   $\lvert g_i\rvert$   &   $\lvert g_i\rvert$   &  $\lvert g_i\rvert$  &   & $\lvert g_i\rvert$        \\
$\pi \Sigma$            &  $3.00$     &  $1.59$   &   $0.26$  & $\pi \Lambda$   &  $-$  \\
$\bar{K}N$              &  $2.41$     &  $3.14$   &   $0.64$   & $\pi \Sigma$     &  $-$   \\
$\eta \Lambda$       &  $0.71$     &  $1.21$   &   $1.78$   & $\bar{K}N$       &   $-$   \\
$K \Xi$                    &  $1.86$    &  $2.14$    &  $4.50$    &  $\eta \Sigma$  & $-$   \\
                               &           &           &            & $K \Xi$  &  $-$    \\
\hline
\hline
\multicolumn{6}{c}{ {\bf s-wave (old)} } \\
\hline
  {\bf $J^P=\frac{1}{2}^-$}  &\multicolumn{3}{|c}{{\bf$(I,S)=(0,-1)$}} &  \multicolumn{2}{|c}{{\bf$(I,S)=(1,-1)$}}  \\
  [0.5mm]
\hline
   &    \multicolumn{2}{c|}{$\Lambda(1405)$}    &  $\Lambda(1670)$ &  & $\Sigma^*$ \\
  &     &     &  &   &   \\
  [-4.5mm]
$M\;\rm[MeV]$             &  $1419^{+16}_{-22}$ &  $1420^{+15}_{-21}$   & $1675^{+10}_{-11}$ & $M\;\rm[MeV]$  & $1701^{+16}_{-1}$  \\
$\Gamma\;\rm[MeV]$   &  $142^{+48}_{-62}$   &  $54^{+36}_{-22}$   & $62^{+8}_{-14}$ & $\Gamma\;\rm[MeV]$ & $340^{+4}_{-14}$ \\
\hline
                                &   $\lvert g_i\rvert$   &   $\lvert g_i\rvert$   &  $\lvert g_i\rvert$  &   & $\lvert g_i\rvert$       \\
$\pi \Sigma$            &  $3.40$     &  $2.31$   &   $0.47$  & $\pi \Lambda$   &  $1.96$  \\
$\bar{K}N$              &  $2.98$     &  $3.51$   &   $0.59$   & $\pi \Sigma$     &  $0.47$   \\
$\eta \Lambda$       &  $1.10$     &  $1.26$   &   $1.74$   & $\bar{K}N$       &   $1.21$   \\
$K \Xi$                    &  $0.65$    &  $0.36$    &  $3.71$    &  $\eta \Sigma$  & $0.36$   \\
                               &                 &                  &                & $K \Xi$              &  $0.98$    \\
\hline
\hline
\end{tabular}
\end{table*}
Finally, we analyze the pole content of the scattering amplitudes derived from the {\bf s+p-waves} and {\bf s-wave} models and compare with the corresponding pole content from the {\bf s-wave (old)}
model. The first thing to be mentioned is that we cannot dynamically generate any state neither with $J^P={\frac{1}{2}}^+$ nor with $J^P={\frac{3}{2}}^+$ from the {\bf s+p-waves} model, in spite of the existence of some $\Lambda^*$ and $\Sigma^*$ resonances with these quantum numbers in the range covered by the higher-energy inelastic channels. According to the PDG compilation~\cite{PDG}, these resonances decay mostly into $\bar{K}N$, $\pi \Sigma$, $\pi \Lambda$ or to more exotic two-{} and three-body products. Thus, the inclusion of higher-energy experimental data from the $K^- p \to \bar{K}N$, $\pi \Sigma$, and $\pi \Lambda$ processes in future fits could favor the appearance of poles in the scattering amplitude, since there are clear structures reflecting resonant signals in the corresponding cross sections.

The results of the pole position of the resonances with $J^P={\frac{1}{2}}^-$, together with their couplings to the different channels, are compiled in Table~\ref{tab:spectroscopy} for the {\bf s+p-waves} (top), the {\bf s-wave} (middle) and the {\bf s-wave (old)} (bottom) models.

In the $I=0$ channel, all models generate the double-pole structure of $\Lambda(1405)$ and a pole corresponding to $\Lambda(1670)$. As can be appreciated from the panel on the top, the broader pole of $\Lambda(1405)$ has been shifted $20$~MeV towards lower energies compared to the corresponding resonance found in the {\bf s-wave (old)} amplitudes. Despite the location of this first pole is not well established, given the scattered positions found in the literature, the new position of this pole is in better agreement with most of the studies devoted to this topic (see e.g.~\cite{Cieply:2016jby,Feijoo:2018den}). This is also the case for the corresponding pole in the {\bf s-wave} model of the middle panel, but in this case it is shifted downwards by 55~MeV. The widths of these broad poles behave in a contrary way, while that of {\bf s+p-waves}
model
gets narrower, compared to that of the {\bf s-wave (old)} model, the one obtained from the {\bf s-wave}
model
is even larger. All these broad poles couple most notably to $\pi \Sigma$ and $\bar{K}N$ states, the coupling to $\pi \Sigma$ being slightly higher. All models produce a higher energy and narrower $\Lambda(1405)$ state. While the position in energy is similar for the three models, the pole found by the {\bf s-wave} model is narrower by more than $10$~MeV compared to those of found by the two other models. The pattern of the couplings for the narrow $\Lambda(1405)$ in the three models is in accordance to most of the works in literature, meaning that this pole couples mostly to the $\bar{K}N$ channel but has a sizable coupling to the $\pi \Sigma$ one.
As for the $\Lambda(1670)$ resonance, the most remarkable fact is that the {\bf s+p-waves} model produces a width that is half of that obtained by the other two models and in much better agreement with the value quoted by the PDG~\cite{PDG}.

Concerning the $I=1$ sector, the {\bf s-wave} model does not generate any state, whereas the {\bf s+p-waves} model produces a wide resonance about 100~MeV below in energy and substantially wider than that found by the {\bf s-wave (old)} model. These states can hardly be identified with any of the $\Sigma$ resonances listed in the PDG with $J^P=1/2^-$~\cite{PDG}. For instance, the observed 1-star $\Sigma(1620)$ resonance is substantially narrower and has been seen to decay into $\pi \Sigma$ states with twice as much probability than into $\pi\Lambda$ states, a decay pattern that cannot be reproduced by neither of the $I=1$ states listed in Table~\ref{tab:spectroscopy} according to the size of their couplings to the different meson-baryon channels. The identification with the $\Sigma(1750)$ resonance is also discarded, as it has been seen to decay more strongly into $\eta\Sigma$ states and, with a somewhat smaller branching ratio, into $\pi\Lambda$ and $\pi\Sigma$. We note, however, that the large width of the $I=1$ states found here hinders their experimental identification in reactions where they might be produced.



\section{Conclusions}
\label{sec:conclusions}

We have performed a new study of the meson--baryon interaction in the $S=-1$ sector including both s-{} and p-waves,
aimed
at improving our knowledge about the NLO terms of the chiral SU(3) Lagrangian. We pay special attention to the processes that are very sensitive to these
subleading
terms of the Lagrangian, such as the $K^- p\to K^+\Xi^-, K^0\Xi^0, \eta \Lambda, \eta \Sigma^0$ reactions.
%
This work presents a detailed derivation of the formalism needed to obtain the higher partial waves of the scattering amplitudes, beyond the s-wave component usually considered in the literature.

Our previous models evolved in the direction of implementing systematic improvement to obtain a more reliable determination of the NLO terms of the Lagrangian, namely the $b_i$ and $d_i$ LECs, culminating into the {\bf s-wave (old)} set obtained in~\cite{Feijoo:2018den}. In the present work we have first improved upon the later model by incorporating the novel $g$-{} and $h$-terms from the NLO interaction kernel. This has led to the new {\bf s-wave} set of parameters, which reproduces experimental data very well and
achieves
very low
$\chi^2_{d.o.f.}$ of 0.77. However, we have observed that the LECs of this {\bf s-wave} model, both the LO and NLO ones, are qualitatively very different from our earlier estimations.

We have then checked how stable is this new {\bf s-wave} set by incorporating into the model the p-wave contributions, which strongly depend on the novel $g$-{} and $h$-terms but also on the rest of NLO parameters. With respect to the {\bf s-wave} results, the {\bf s+p-waves} model also produces a very good fit, although with a slightly worse $\chi^2_{d.o.f.}$ value of 0.86 and a rather different set of LECs. The latter fact is an indication of the clear interconnection between the higher partial waves and the higher orders of the chiral expansion, as both contributions acquire importance at higher energies. It is interesting to note, however, that the size of the LECs obtained with the {\bf s+p-waves}
model is smaller and closer to that of our earlier {\bf s-wave (old)} model. This makes us believe that the {\bf s+p-waves} model developed in the present work constitutes a very good starting point for subsequent implementations of higher partial waves. In this work, we have
shown that a notable fraction of the cross sections of the $K^- p\to K^+\Xi^-, K^0\Xi^0, \eta \Lambda, \eta \Sigma^0$ reactions comes indeed from the novel p-wave contributions, thereby making it clear that higher partial waves cannot be considered as minor corrections in these processes.

The main conclusion of this study is that one does need to incorporate higher partial-wave contributions to the scattering amplitudes in order to properly describe the $K^- p$ inelastic reactions opening up at higher energies. This is in contrast to our previous works, where the experimental data of such processes was reasonably reproduced employing pure s-wave scattering amplitudes. In other words, our previous models have effectively overestimated the lowest partial-wave contributions, thereby masking the physics of higher partial waves behind the values of the NLO parameters. As for the relevance of the new $g$-{} and $h$-terms considered in this work, compared to the rest of NLO contributions, we cannot say anything conclusive, as we have obtained very different values of the parameters in the two new fits presented here. What seems to be clear is that they provide a non-negligible contribution to the scattering amplitudes, especially far enough from thresholds, because of their strong dependence on momenta. Therefore, it is natural to consider the $g$-{} and $h$-terms when implementing higher partial waves in the scattering amplitudes, since both contributions acquire a relevant role at higher energies.

%
As a final remark we would like to comment that the {\bf s+p-waves} model
is the natural continuation of our previous ones~\cite{Feijoo:2015yja, Ramos:2016odk, Feijoo:2018den}. Since it employs the same set of data points in the fit, it provides a clear quantification of the effect of the new ingredients, namely the $g$-{} and $h$-terms and the p-waves.
This model should then be considered as a promising reference point upon which additional partial-wave contributions,
in particular the d-waves, need to be added. However, this challenging study, which is already in progress, requires including more experimental data points of
total and differential cross sections within a
wider
range of energies in the fits.
As can be clearly seen in Table~\ref{tab:outputs_fits}, the determination of LECs becomes an extensive nonlinear optimization problem where the relatively large number of parameters makes it challenging to determine their optimal values. Thus, it is also crucial to establish a credible program to quantify the magnitude of statistical and systematic uncertainties of the LECs and to study their propagation into experimental observables. This can be achieved by employing advanced mathematical optimization tools, as introduced e.g.\ in~\cite{NavarroPerez:2012qf,Carlsson:2015vda}.
We believe that this is the proper procedure to have a realistic determination of the NLO parameters of the chiral SU(3) Lagrangian.

\vspace{6pt}




\authorcontributions{All authors contributed on an equal footing to the research presented in this work. All authors have read and
agreed to the published version of the manuscript.
}

\funding{V.M.\ and A.R.\ acknowledge support from the Ministerio de Ciencia, Innovaci\'on y Universidades under the project CEX2019-000918-M of ICCUB (Unidad de Excelencia ``Mar\' ia de Maeztu''), and from the Ministerio de Econom\' ia y Competitividad, with additional European Regional Development Funds, under the contract FIS2017-87534-P. A.F.\ acknowledges support from the from the Spanish Ministerio de Ciencia, Innovaci\'on y Universidades under the project PID2020-112777GB-I00, from the Ministerio de Econom\' ia y Competitividad and the European Regional Development Fund, under contract FIS2017-84038-C2-1-P, and from the Generalitat Valenciana under contract PROMETEO/2020/023. V.M., A.R.\ and A.F.\ also acknowledge support from the EU STRONG-2020 project under the program H2020-INFRAIA-2018-1, grant agreement no.\ 824093. The work of D.G.\ was supported by the Czech Science Foundation, GA\v{C}R grant no.\ 19-19640S.}


\conflictsofinterest{The authors declare no conflict of interest.}
\vskip 3cm


\appendixtitles{no} 
\appendix
\vspace*{-2cm}
\section{}
\label{ghcoeffs}
\unskip
\begin{sidewaystable*}[ht!]
  \caption{The $g_{ij}$ and $h_{ij}$ coefficients of Eq.~(\ref{Kernel_NLO}).}
  \label{tab_g_h_coef}
\centering
\scalebox{0.6}{
\begin{tabular}{|l|cccccccccc|}
\multicolumn{11}{c}{{\rm $g_{ij}$ coefficients}} \\[2.5mm]
\hline
 & & & & & & & & & & \\[-2.5mm]
  &{\bf $K^-p$}&{\bf $\bar{K}^0n$}&{\bf $\pi^0\Lambda$}& {\bf $\pi^0\Sigma^0$}&{\bf $\eta\Lambda$}&{\bf $\eta\Sigma^0$}&{\bf $\pi^+\Sigma^-$}&{\bf $\pi^-\Sigma^+$}&{\bf $K^+\Xi^-$}&{\bf $K^0\Xi^0$}  \\
 \hline
& & & & & & & & & & \\[-2.5mm]
{\bf $K^-p$} & $2(g_2+g_4)$ & $g_1+g_2$ & $-\frac{\sqrt{3}}{2}(g_1+g_2)$ & $-\frac{1}{2}(g_1+g_2)$  & $\frac{1}{2}(g_1-3g_2)$  & $\frac{1}{2\sqrt{3}}(g_1-3g_2)$  & $-2g_2$& $g_2-g_1$  & $-4g_2$  & $-2g_2$  \\ [2mm]
{\bf $\bar{K}^0n$}   &  &$2(g_2+g_4)$ &$\frac{\sqrt{3}}{2}(g_1+g_2)$  & $-\frac{1}{2}(g_1+g_2)$ &  $\frac{1}{2}(g_1-3g_2)$  & $-\frac{1}{2\sqrt{3}}(g_1-3g_2)$   & $g_2-g_1$   & $-2g_2$ & $-2g_2$  &  $-4g_2$    \\ [2mm]
{\bf $\pi^0\Lambda$} &    &    &  $2g_4$  &  $0$ & $0$ &  $0$ & $0$ &  $0$ & $\frac{\sqrt{3}}{2}(g_1-g_2)$  & $\frac{\sqrt{3}}{2}(g_2-g_1)$ \\ [2mm]
{\bf $\pi^0\Sigma^0$}&    &    &      & $2g_4$ & $0$  & $0$ & $-2g_2$ & $-2g_2$ & $\frac{1}{2}(g_1-g_2)$ & $\frac{1}{2}(g_1-g_2)$  \\ [2mm]
{\bf $\eta\Lambda$}  &     &    &      &       & $2g_4$  & $0$  & $0$ & $0$ & $-\frac{1}{2}(g_1+3g_2)$ & $-\frac{1}{2}(g_1+3g_2)$ \\ [2mm]
{\bf $\eta\Sigma^0$} &    &     &     &     &   & $2g_4$ & $\frac{2}{\sqrt{3}}g_1$ &$-\frac{2}{\sqrt{3}}g_1$ &  $-\frac{1}{2\sqrt{3}}(g_1+3g_2)$ & $\frac{1}{2\sqrt{3}}(g_1+3g_2)$   \\ [2mm]
{\bf $\pi^+\Sigma^-$}&    &     &     &     &       &   & $2(g_2+g_4)$ & $-4g_2$ & $g_1+g_2$ & $-2g_2$ \\ [2mm]
{\bf $\pi^-\Sigma^+$}&    &    &     &      &     &     &     & $2(g_2+g_4)$ & $-2g_2$ & $g_1+g_2$  \\ [2mm]
{\bf $K^+\Xi^-$}     &     &     &     &       &       &        &     &     & $2(g_2+g_4)$  &  $g_2-g_1$ \\ [2mm]
{\bf $K^0\Xi^0$}     &     &     &               &       &       &               &     &     &    & $2(g_2+g_4)$  \\ [2mm]
\hline
\multicolumn{11}{c}{} \\[15mm]
\multicolumn{11}{c}{{\rm $h_{ij}$ coefficients}} \\[2.5mm]
\hline
 & & & & & & & & & & \\[-2.5mm]
  &{\bf $K^-p$}&{\bf $\bar{K}^0n$}&{\bf $\pi^0\Lambda$}& {\bf $\pi^0\Sigma^0$}&{\bf $\eta\Lambda$}&{\bf $\eta\Sigma^0$}&{\bf $\pi^+\Sigma^-$}&{\bf $\pi^-\Sigma^+$}&{\bf $K^+\Xi^-$}&{\bf $K^0\Xi^0$}  \\
 \hline
& & & & & & & & & & \\[-2.5mm]

{\bf $K^-p$} & $-h_1-h_2-h_3-h_4$ & $-h_2-h_3-h_4$ & $\frac{1}{2\sqrt{3}}(-h_1+3h_2+h_3)$  & $\frac{1}{2}(-h_1+h_2-h_3-2h_4)$ & $\frac{1}{6}(h_1+3h_2-7h_3-6h_4)$   & $\frac{1}{2\sqrt{3}}(h_1+h_2-h_3)$ & $h_2-h_3-h_4$ & $-h_1-h_4$ & $2(h_2-h_3-h_4)$ & $h_2-h_3-h_4$ \\ [2mm]
{\bf $\bar{K}^0n$}   &    & $-h_1-h_2-h_3-h_4$ & $\frac{1}{2\sqrt{3}}(h_1-3h_2-h_3)$ & $\frac{1}{2}(-h_1+h_2-h_3-2h_4)$ & $\frac{1}{6}(h_1+3h_2-7h_3-6h_4)$ & $\frac{1}{2\sqrt{3}}(-h_1-h_2+h_3)$  & $-h_1-h_4$ & $h_2-h_3-h_4$ & $h_2-h_3-h_4$ & $2(h_2-h_3-h_4)$  \\ [2mm]
{\bf $\pi^0\Lambda$} &    &    &  $-\frac{1}{3}(h_1+2h_3)$ & $0$ & $0$ & $-\frac{1}{3}(h_1+2h_3+3h_4)$  & $0$  & $0$ & $\frac{1}{\sqrt{3}}(h_1-h_3)$ &$\frac{1}{\sqrt{3}}(h_3-h_1)$ \\ [2mm]
{\bf $\pi^0\Sigma^0$}&    &    &      &  $-h_1-2h_3-2h_4$ &  $-\frac{1}{3}(h_1+2h_3+3h_4)$  & $0$ & $h_2-h_3-h_4$ & $h_2-h_3-h_4$ & $-h_3-h_4$ &  $-h_3-h_4$ \\ [2mm]
{\bf $\eta\Lambda$}  &     &    &      &   & $-h_1-2h_3-2h_4$  & $0$ & $-\frac{1}{3}(h_1+2h_3+3h_4)$  & $-\frac{1}{3}(h_1+2h_3+3h_4)$ & $\frac{1}{3}(-h_1+3h_2-2h_3-3h_4)$ &  $\frac{1}{3}(-h_1+3h_2-2h_3-3h_4)$    \\ [2mm]
{\bf $\eta\Sigma^0$} &    &     &     &     &   & $\frac{1}{3}(-h_1-2h_3)$ & $\frac{1}{\sqrt{3}}(h_1-h_2-h_3)$ & $-\frac{1}{\sqrt{3}}(h_1-h_2-h_3)$ &  $\frac{1}{\sqrt{3}}h_2$ &   $-\frac{1}{\sqrt{3}}h_2$ \\ [2mm]
{\bf $\pi^+\Sigma^-$}&    &     &     &     &       &   & $-h_1-h_2-h_3-h_4$ & $2(h_2-h_3-h_4)$  &$-h_2-h_3-h_4$ & $h_2-h_3-h_4$   \\ [2mm]
{\bf $\pi^-\Sigma^+$}&    &    &     &      &     &     &     & $-h_1-h_2-h_3-h_4$ & $h_2-h_3-h_4$ &  $-h_2-h_3-h_4$ \\ [2mm]
{\bf $K^+\Xi^-$}     &     &     &     &       &       &        &     &     & $-h_1-h_2-h_3-h_4$ & $-h_1-h_4$   \\ [2mm]
{\bf $K^0\Xi^0$}     &     &     &               &       &       &               &     &     &    & $-h_1-h_2-h_3-h_4$  \\ [2mm]

\hline
\end{tabular}
}
\end{sidewaystable*}




\reftitle{References}


\begin{thebibliography}{999}

\bibitem{Gasser:1983yg}
Gasser,~J.; Leutwyler,~H.
Chiral Perturbation Theory to One Loop,
{\em Annals Phys.} {\bf 1984}, {\em 158}, 142.

\bibitem{Guo:2017jvc}
Guo, F.~K.; Hanhart, C.; Mei\ss{}ner, U.~G.; Wang, Q.; Zhao, Q.; Zou, B.~S.
Hadronic molecules,
{\em  Rev. Mod. Phys.} {\bf 2018}, {\em 90}, 015004.

\bibitem{Dalitz:1959dn}
Dalitz, R.~H.; Tuan, S.~F.
A possible resonant state in pion-hyperon scattering,
{\em Phys. Rev. Lett. } {\bf 1959}, {\em 2}, 425-428.
doi:10.1103/PhysRevLett.2.425

\bibitem{Dalitz:1960du}
Dalitz, R.~H.; Tuan, S.~F.
The phenomenological description of -K -nucleon reaction processes,
{\em Annals Phys. } {\bf 1960}, {\em 10}, 307-351.
doi:10.1016/0003-4916(60)90001-4

\bibitem{Kaiser:1995eg}
Kaiser, N.; Siegel, P.~B.; Weise, W.
Chiral dynamics and the low-energy kaon - nucleon interaction,
{\em Nucl. Phys. A} {\bf 1995}, {\em 594}, 325-345.

\bibitem{Oset:1997it}
Oset, E.; Ramos, A. Nonperturbative chiral approach to s-wave $\bar{K}N$ interactions, {\em Nucl. Phys. A} {\bf 1998}, {\em 635}, 99-120.


\bibitem{Oller:2000fj}
Oller, J.~A.; Meissner, U.~G.
Chiral dynamics in the presence of bound states: Kaon nucleon interactions revisited,
{\em Phys. Lett. B} {\bf 2001}, {\em 500}, 263-272.

\bibitem{Lutz:2001yb}
Lutz, M.~F.~M.; Kolomeitsev, E.~E. Relativistic chiral SU(3) symmetry, large N(c) sum rules and meson baryon scattering, {\em Nucl. Phys. A} {\bf 2002}, {\em 700}, 193-308.

\bibitem{GarciaRecio:2002td}
Garcia-Recio, C.; Nieves, J.; Ruiz Arriola, E.; Vicente Vacas, M.~J. $S =-1$ meson baryon unitarized coupled channel chiral perturbation theory and the S(01) $\Lambda(1405)$ and $\Lambda(1670)$ resonances, {\em Phys. Rev. D} {\bf 2003}, {\em 67}, 076009.

\bibitem{Borasoy:2005ie}
Borasoy, B.; Nissler, R.; Weise, W. Chiral dynamics of kaon-nucleon interactions, revisited, {\em Eur. Phys. J. A} {\bf 2005}, {\em 25}, 79-96.

\bibitem{Jido:2003cb}
Jido, D.; Oller, J.~A.; Oset, E.; Ramos, A.; Meissner, U.~G.
Chiral dynamics of the two Lambda(1405) states,''
{\em Nucl. Phys. A} {\bf 2003}, {\em 725}, 181-200.


\bibitem{Magas:2005vu}
Magas, V.~K.; Oset, E.; Ramos, A.
Evidence for the two pole structure of the Lambda(1405) resonance,
{\em Phys. Rev. Lett.} {\bf 2005}, {\em 95}, 052301.



\bibitem{Meissner:2020khl}
Mei\ss{}ner, U.~G.
Two-pole structures in QCD: Facts, not fantasy!,
{\em Symmetry} {\bf 2020}, {\em 12}, 981.

\bibitem{Zychor:2007gf}
Zychor, I.; Buscher, M.; Hartmann, M.; Kacharava, A.; Keshelashvili, I.; Khoukaz, A.; Kleber, V.; Koptev, V.; Maeda, Y.; Mersmann, T.; \textit{et al.}
~Shape of the Lambda(1405) hyperon measured through its Sigma0 pi0 Decay,
{\em  Phys. Lett. B } {\bf 2008}, {\em 660}, 167-171.

\bibitem{HADES:2012csk}
Agakishiev, G. \textit{et al.} [HADES],
Baryonic resonances close to the $\bar{K}N$ threshold: the case of $\Lambda$(1405) in {\em Phys. Rev. C} {\bf 2013}, {\em 87}, 025201.


\bibitem{Niiyama:2008rt}
Niiyama, M.; Fujimura, H.; Ahn, D.~S.; Ahn, J.~K.; Ajimura, S.; Bhang, H.~C.; Chang, T.~H.; Chang, W.~C.; Chen, J.~Y.; Date, S. \textit{et al.}
Photoproduction of Lambda(1405) and Sigma0(1385) on the proton at E(gamma) = 1.5-2.4-GeV,
{\em Phys. Rev. C} {\bf 2008}, {\em 78}, 035202.

\bibitem{CLAS:2013rjt}
Moriya, K.; \textit{et al.} [CLAS],
Measurement of the \ensuremath{\Sigma}\ensuremath{\pi} photoproduction line shapes near the \ensuremath{\Lambda}(1405),
{\em Phys. Rev. C} {\bf 2013}, {\em 87}, 035206.

\bibitem{CLAS:2013rxx}
Moriya, K.; \textit{et al.} [CLAS],
Differential Photoproduction Cross Sections of the $\Sigma^0(1385)$, $\Lambda(1405)$, and $\Lambda(1520)$,
{\em Phys. Rev. C} {\bf 2013}, {\em 88}, 045201.

\bibitem{CLAS:2013zie}
Lu, H.~Y.; \textit{et al.} [CLAS],
First Observation of the $\Lambda(1405)$ Line Shape in Electroproduction,
{\em Phys. Rev. C} {\bf 2013}, {\em 88}, 045202.

\bibitem{SIDD}
Bazzi, M.; {\it et al.} (SIDDHARTA) A New Measurement of Kaonic Hydrogen X-rays, {\em Phys. Lett. B} {\bf 2011}, {\em 704}, 113-117.


\bibitem{GO}
Guo, Z.~H.; Oller, J.~A. Meson-baryon reactions with strangeness -1 within a chiral framework, {\em Phys. Rev. C} {\bf 2013}, {\em 87}, 035202.

\bibitem{IHW}
Ikeda, Y.; Hyodo, T.; Weise, W. Chiral SU(3) theory of antikaon-nucleon interactions with improved threshold constraints, {\em Nucl. Phys. A} {\bf 2012}, {\em 881}, 98-114.

\bibitem{Roca:2013cca}
Roca, L.; Oset, E.
Isospin 0 and 1 resonances from $\pi \Sigma$ photoproduction data,
{\em  Phys. Rev. C} {\bf 2013}, {\em 88}, 055206.

\bibitem{Mai:2014xna}
Mai, M.; Mei\ss{}ner, U.~G. Constraints on the chiral unitary $\bar KN$ amplitude from $\pi\Sigma K^+$ photoproduction data, {\em Eur. Phys. J. A} {\bf 2015}, {\em 51}, 30.

\bibitem{Cieply:2016jby}
Ciepl\'y, A.; Mai, M.; Mei\ss{}ner, U.~G.; Smejkal, J. On the pole content of coupled channels chiral approaches used for the $\bar{K}N$ system, {\em Nucl. Phys. A} {\bf 2016}, {\em 954}, 17-40.

\bibitem{Feijoo:2015yja}
Feijoo, A.; Magas, V.; Ramos, A. The $\bar{K} N \rightarrow K \Xi$ reaction in coupled channel chiral models up to next-to-leading order,
{\em Phys. Rev. C} {\bf 2015}, {\em 92}, 015206.

\bibitem{Ramos:2016odk}
Ramos, A.; Feijoo, A.; Magas, V.~K. The chiral S = \ensuremath{-}1 meson\textendash{}baryon interaction with new constraints on the NLO contributions, {\em Nucl. Phys. A} {\bf 2016}, {\em 954}, 58-74.

\bibitem{Feijoo:2018den}
Feijoo, A.; Magas, V.; Ramos, A. $S$=\ensuremath{-}1 meson-baryon interaction and the role of isospin filtering processes,
{\em Phys. Rev. C} {\bf 2019}, {\em 99}, 035211.

\bibitem{Feijoo:2015cca}
Feijoo, A.; Magas, V.; Ramos, A.; Oset, E.
$\Lambda_b \to J/\psi ~ K ~ \Xi$ decay and the higher order chiral terms of the meson baryon interaction,
{\em Phys. Rev. D} {\bf 2015 }, {\em 92}, no.7, 076015
[erratum: {\em Phys. Rev. D } {\bf 2017}, {\em 95}, no.3, 039905].

\bibitem{Garcia-Recio:2000seh}
Garcia-Recio, C.; Oset, E.; Ramos, A.; Nieves, J.
Nonlocalities and Fermi motion corrections in K- atoms,
{\em Nucl. Phys. A} {\bf 2002}, {\em 703}, 271-294.

\bibitem{Cieply:2011yz}
A.~Cieply, E.~Friedman, A.~Gal, D.~Gazda and J.~Mares,
Phys. Lett. B \textbf{702}, 402-407 (2011)
doi:10.1016/j.physletb.2011.07.040
[arXiv:1102.4515 [nucl-th]].

\bibitem{Hrtankova:2017zxw}
Hrt\'ankov\'a, J.; Mare\v{s}, J.
$K^-$- nuclear states: Binding energies and widths,
{\em Phys. Rev. C} {\bf 2017}, {\em 96}, 015205.

\bibitem{Hrtankova:2019jky}
Hrt\'ankov\'a, J.; Ramos, \`A.
Single- and two-nucleon antikaon absorption in nuclear matter with chiral meson-baryon interactions,
{\em Phys. Rev. C} {\bf 2020}, {\em 101}, 035204.


\bibitem{CaroRamon:1999jf}
Caro Ramon, J.; Kaiser, N.; Wetzel, S.; Weise, W. Chiral SU(3) dynamics with coupled channels: Inclusion of P wave multipoles, {\em Nucl. Phys. A} {\bf 2000}, {\em 672}, 249-269.

\bibitem{Aoki:2018wug}
Aoki, K.; Jido, D. $KN$ scattering amplitude revisited in a chiral unitary approach and a possible broad resonance in $S =+1$ channel, {\em PTEP} {\bf 2019}, {\em 2019}, 013D01.


\bibitem{Jido:2002zk}
Jido, D.; Oset, E.; Ramos, A. Chiral dynamics of p wave in $K^- p$ and coupled states, {\em Phys. Rev. C} {\bf 2002}, {\em 66}, 055203.

\bibitem{Cieply:2015pwa}
Ciepl\'y, A.; Krej\v{c}i\v{r}\'\i{}k, V.
Effective model for in-medium $\bar{K}N$ interactions including the $L=1$ partial wave,
{\em Nucl. Phys. A} {\bf 2015},  {\em 940}, 311-330.


\bibitem{Sadasivan:2018jig}
Sadasivan, D.; Mai, M.; D\"oring, M.
S- and p-wave structure of $S=-1$ meson-baryon scattering in the resonance region,
{\em Phys. Lett. B} {\bf 2019}, {\em 789}, 329-335.


\bibitem{Hyodo:2011ur}
Hyodo, T.; Jido, D. The nature of the $\Lambda(1405)$ resonance in chiral dynamics, {\em Prog. Part. Nucl. Phys.} {\bf 2012}, {\em 67}, 55-98.

\bibitem{Scherer:2002tk}
Scherer, S. Introduction to chiral perturbation theory, {\em Adv. Nucl. Phys.} {\bf 2003}, {\em 27}, 277.


\bibitem{Gasser:1990ce}
Gasser, J.; Leutwyler, H.; Sainio, M.~E. Sigma term update, {\em Phys.\ Lett.\ B} {\bf 1991}, {\em 253}, 252.

 \bibitem{Koch:1985bn}
Koch, R. A Calculation of Low-Energy $\pi N$ Partial Waves Based on Fixed t Analyticity, {\em Nucl.\ Phys.\ A} {\bf 1986}, {\em 448}, 707.

 \bibitem{Dover:1982zh}
Dover, C.~B.; Walker, G.~E. The Interaction Of Kaons With Nucleons And Nuclei, {\em Phys.\ Rept.} {\bf 1982}, {\em 89}, 1.


\bibitem{Bernard:1995dp}
Bernard, V.; Kaiser, N.; Meissner, U.~G. Chiral dynamics in nucleons and nuclei, {\em Int. J. Mod. Phys. E} {\bf 1995}, {\em 4}, 193-346.

\bibitem{Fettes:2000gb}
Fettes, N.; Meissner, U.~G.; Mojzis, M.; Steininger, S.; The Chiral effective pion nucleon Lagrangian of order p**4, {\em Annals Phys.} {\bf 2000}, {\em 283}, 273-302; Erratum: {\em Annals Phys.} {\bf 2001}, {\em 288}, 249-250.



\bibitem{Frink:2004ic}
Frink, M.; Meissner, U.~G. Chiral extrapolations of baryon masses for unquenched three flavor lattice simulations, {\em JHEP} {\bf 2004}, {\em 07}, 028.

\bibitem{Oller:2006yh}
Oller, J.~A.; Verbeni, M.; Prades, J. Meson-baryon effective chiral lagrangians to O(q**3), {\em JHEP} {\bf 2006}, {\em 09}, 079.






\bibitem{exp_1}
Kim, J.~K. Low-Energy $K^- p$ Interaction of the 1405-MeV $Y^*_0$ Resonance as $\bar{K}N$ Bound State, {\em Phys. Rev. Lett.} {\bf 1965}, {\em 14}, 29.

\bibitem{exp_2}
Mast, T.~S.; Alston-Garnjost, M.; Bangerter, R.~O.; Barbaro-Galtieri, A.~S.; Solmitz, F.~T.; Tripp, R.~D. Elastic, Charge Exchange, and Total $K^- p$ Cross-Sections in the Momentum Range 220-MeV/c to 470-MeV/c, {\em Phys. Rev. D} {\bf 1976}, {\em 14}, 13.

\bibitem{exp_3}
Bangerter, R.~O.; Alston-Garnjost, M.; Barbaro-Galtieri, A.; Mast, T.~S.; Solmitz, F.~T.; Tripp, R.~D. Reactions $K^- p \to \Sigma^- \pi^+$ and $K^- p \to \Sigma^+ \pi^-$ in the Momentum Range From 220-\{MeV\}/c to 470-\{MeV\}/c, {\em Phys. Rev. D} {\bf 1981}, {\em 23}, 1484.

\bibitem{exp_4}
Ciborowski, J.; Gwizdz, J.; Kielczewska, D.; Nowak, R.~J.; Rondio, E.; Zakrzewski, J.~A.; Goossens, M.; Wilquet, G.; Bedford, N.~H.; Evans, D.; {\it et al.} Kaon scattering and charged $\Sigma$ hyperon production in $K^- p$ interactions below 300-MeV/c, {\em J. Phys. G} {\bf 1982}, {\em 8}, 13-32.

\bibitem{exp_5}
Burgun, G.; Meyer, J.; Pauli, E.; Tallini, B.; Vrana, J.; De Bellefon, A.; Berthon, A.; Rangan, K.~L.; Beaney, J.; Deen, S.~M.; {\it et al.} Resonance formation in the reactions
$K^- + p \to K^+ + \Xi^-$ and $K^- + p \to K^0 + \Xi^0$ in the mass region from 1915 to 2168~MeV, {\em Nucl. Phys. B} {\bf 1968}, {\em 8}, 447-459.

\bibitem{exp_6}
Carlson, J.~R.; Davis, H.~F.; Jauch, D.~E.; Sossong, N.~D.; Ellsworth, R. Measurement of neutral cascade production from negative-kaon-hydrogen at 1.8~GeV/c, $K^- p \to K^0 \Xi^0$, {\em Phys. Rev. D} {\bf 1973}, {\em 7}, 2533-2537.

\bibitem{exp_7}
Dauber, P.~M.; Berge, J.~P.; Hubbard, J.~R.; Merrill, D.~W.; Muller, R.~A. Production and decay of cascade hyperons, {\em Phys. Rev.} {\bf 1969}, {\em 179}, 1262-1285.

\bibitem{exp_8}
Haque, M.; {\it et al.} (Birmingham-Glasgow-London(I.C.)-Oxford-Rutherford) Reactions $K^- p \to$ Hyperon + Meson at 3.5~GeV/c, {\em Phys. Rev.} {\bf 1966}, {\em 152}, 1148-1161.

\bibitem{exp_9}
London, G.~W.; Rau, R.~R.; Samios, N.~P.; Yamamoto, S.~S.; Goldberg, M.; Lichtman, S.; Prime, M.; Leitner, J. $K^- p$ Interaction at 2. Be-24Vc, {\em Phys. Rev.} {\bf 1966}, {\em 143}, 1034-1091.

\bibitem{exp_10}
Trippe, T.~G.; Schlein, P.~E. Partial-Wave Analysis of $K^- p \to \Xi^- K^+$ at 2~GeVc, {\em Phys. Rev.} {\bf 1967}, {\em 158}, 1334-1337.

\bibitem{exp_11}
Trower, W.~P.; Ficenec, J.~R.; Hulsizer, R.~I.; Lathrop, J.; Snyder, J.~N.; Swanson, W.~P. Some Two-Body Final States of $K^- p$ Interactions at 1.33~GeVc, {\em Phys. Rev.} {\bf 1968}, {\em 170}, 1207-1222.

\bibitem{Starostin}
Starostin, A.; Nefkens, B.~M.~K.; Berger, E.; Clajus, M.; Marusic, A.; McDonald, S.; Phaisangittisakul, N.; Prakhov, S.; Price, J.~W.; Pulver, M.; {\it et al.} (Crystal Ball Collaboration) Measurement of $K^- p \to \eta \Lambda $ near threshold, {\em Phys. Rev. C} {\bf 2001}, {\em 64}, 055205.

\bibitem{Baxter}
Baxter, D.~F.; Buckingham, I.~D.; Corbett, I.~F.; Dunn, P.~A.; Emmerson, J.~M.; Garvey, J.; Hart, F.; Hughes, G.; Jones, C.~M.~S.; Maybury, R.; {\it et al.} A study of neutral final states in $K^- p$ interactions in the range from 690 to 934~MeV/c, {\em Nucl. Phys. B} {\bf 1973}, {\em 67}, 125-156.

\bibitem{Jones}
Jones, M.; Levi Setti, R.; Merrill, D.; Tripp, R.~D. $K^- p$ Charge Exchange and Hyperon Production Cross-Sections from 860-MeV/c to 1000-MeV/c, {\em Nucl. Phys. B} {\bf 1975}, {\em 90}, 349-383.

\bibitem{Berthon}
Berthon, A.; Tristram, G.; Vrana, J.; Bacon, T.~C.; Brandstetter, A.~A.; Butterworth, I.; Gopal, G.~P.; Jones, P.~S.; Litchfield, P.~J.; Mandelkern, M.; {\it et al.} Cross-sections for quasi-two-body reactions in $K^- p$ interactions between 1263 and 1843~MeV/c, {\em Nuovo Cim. A} {\bf 1974}, {\em 21}, 146-167.

\bibitem{Jones:1974si}
Jones, M.~D.; A study of the reaction $K^- p \to \Sigma^0 \eta$ near threshold, {\em Nucl. Phys. B} {\bf 1974}, {\em 73}, 141-165.


\bibitem{br_1}
Nowak, R.~J.; Armstrong, J.; Davis, D.~H.; Miller,D.~J.; Tovee, D.~N.; Bertrand, D.; Goossens, M.; Vanhomwegen, G.; Wilquet, G.; Abdullah, M.; {\it et al.} Charged Sigma Hyperon Production by K- Meson Interactions at Rest, {\em Nucl. Phys. B} {\bf 1978}, {\em 139}, 61-71.

\bibitem{br_2}
Tovee, D.~N.; Davis, D.~H.; Simonovic, J.; Bohm, G.; Klabuhn, J.; Wysotzki, F.; Csejthey-Barth, M.; Wickens, J.~H.; Cantwell, T.; Ni Ghogain, C.; {\it et al.} Some properties of the charged sigma hyperons, {\em Nucl. Phys. B} {\bf 1971}, {\em 33}, 493-504.


\bibitem{Meissner:2004jr}
Meissner, U.~G.; Raha, U.; Rusetsky, A. Spectrum and decays of kaonic hydrogen, {\em Eur. Phys. J. C} {\bf 2004}, {\em 35}, 349-357.





\bibitem{Matveev:2019igl}
Matveev, M.; Sarantsev, A.~V.; Nikonov, V.~A.; Anisovich, A.~V.; Thoma, U.; Klempt, E. Hyperon I: Partial-wave amplitudes for K$^{-}$p scattering, {\em Eur. Phys. J. A} {\bf 2019}, {\em 55}, 179.



\bibitem{PDG}
  Zyla, P.~A.; {\it et al.} (Particle Data Group), {\em Prog. Theor. Exp. Phys.} {\bf 2020}, 083C01.


\bibitem{James:1975dr}
F.~James and M.~Roos,
Comput. Phys. Commun. \textbf{10}, 343-367 (1975)
doi:10.1016/0010-4655(75)90039-9

\bibitem{Carlsson:2015vda}
B.~D.~Carlsson, A.~Ekstr\"om, C.~Forss\'en, D.~F.~Str\"omberg, G.~R.~Jansen, O.~Lilja, M.~Lindby, B.~A.~Mattsson and K.~A.~Wendt,
Phys. Rev. X \textbf{6}, no.1, 011019 (2016)
doi:10.1103/PhysRevX.6.011019
[arXiv:1506.02466 [nucl-th]].

\bibitem{NavarroPerez:2012qf}
R.~Navarro P\'erez, J.~E.~Amaro and E.~Ruiz Arriola,
Phys. Lett. B \textbf{724}, 138-143 (2013)
doi:10.1016/j.physletb.2013.05.066
[arXiv:1202.2689 [nucl-th]].

\end{thebibliography}
\end{document}